\documentclass[geosciences,review,accept,oneauthors,pdftex,10pt,a4paper]{mdpi}

\usepackage{upgreek}

\setitemize{parsep=6pt,itemsep=0pt,leftmargin=*,labelsep=5.5mm}
\setenumerate{parsep=6pt,itemsep=0pt,leftmargin=*,labelsep=5.5mm} 
\setlist[description]{itemsep=0mm} 

\usepackage[utf8]{inputenc}
\usepackage{amsmath}
\usepackage{amssymb}
\DeclareUnicodeCharacter{2212}{-}
\usepackage{graphicx}
\usepackage{multirow}
\usepackage{booktabs}
\usepackage{xcolor}
\usepackage{gensymb}
\usepackage{textcomp}
\usepackage{siunitx}

\newcommand*\icarus{Icarus}

\newcommand*\memsai{Mem.~Soc.~Astron.~Ital.}

\newcommand*\nat{Nature}

\usepackage{makecell}

\firstpage{1} 
\articlenumber{x}
\doinum{10.3390/------}
\pubvolume{xx}
\pubyear{2018}
\copyrightyear{2018}
\history{Received: 30 June 2018; Accepted: 31 July 2018; Published: 3 August 2018}

\Title{Multi-Wavelength High-Resolution Spectroscopy for 
 Exoplanet Detection: Motivation, Instrumentation and First Results}

\Author{Serena Benatti} 

\AuthorNames{Serena Benatti}

\address[1]{INAF-Astronomical Observatory of Padova, Vicolo Osservatorio 5, I-35122 Padova, Italy; serena.benatti@inaf.it }

\abstract{Exoplanet research has shown an incessant growth since the first claim of a hot giant planet around a solar-like star in the mid-1990s. Today, the new facilities {are working to spot the first habitable rocky planets} around low-mass stars as a forerunner for the detection of the long-awaited Sun-Earth analog system. All~the achievements in this field would not have been possible without the constant development of the technology and of new methods to detect more and more challenging planets. After the consolidation of a top-level instrumentation for high-resolution spectroscopy in the visible wavelength range, a huge effort is now dedicated to reaching the same precision and accuracy in the near-infrared. Actually, observations in this range present several advantages in the search for exoplanets around M dwarfs, known to be the most favorable targets to detect possible habitable planets. They are also characterized by intense stellar activity, which hampers planet detection, but~its impact on the radial velocity modulation is mitigated in the infrared. Simultaneous observations in the visible and near-infrared ranges appear to be an even more powerful technique since they provide combined and complementary information, also useful for many other exoplanetary science~cases.
}


\begin{document}


\section{Introduction}

The radial velocity (or Doppler) method allowed detecting the first extrasolar planet around a solar-like star, 51 Peg b \cite{1995Natur.378..355M}, opening the era of the quest for planets. The radial velocity (RV) technique is an ``indirect'' method to detect exoplanets since it reveals the reflex motion of the host star due to a hidden companion when both revolve around the common center of mass. The periodical variation of the stellar RV, obtained through the measurement of the Doppler shift of its spectral lines, is dependent on the characteristic of its small-mass companion. The modulation can be represented by a Keplerian function with a period equal to the orbital period of the planet, \textit{P}, and a semi-amplitude, \textit{K}, defined as~follows:
\begin{equation} \label{eqn:K}
K {\rm [m s^{−1}]}=\frac{28.4329{\rm [m s^{−1}]}}{\sqrt{1-e^2}}\frac{m_p \sin i}{M_J}\left( \frac{M_{\star}+m_p}{M_{\odot}}\right)^{-2/3}\left(\frac{P}{{\rm 1[yr]}}\right)^{-1/3},
\end{equation}
where $M_{\star}$ is the stellar mass, $e$ the orbital eccentricity of the planet and $m_p \sin i$ is the lower limit of the planetary mass since the real mass depends on the orbital inclination, \textit{i}, unknown with the RV information alone. {The actual value of the planet mass can be obtained if the planet transits in front of its host star and a proper modeling of the observed transit light curve is performed. Just like the shape of the RV variation allows measuring a set of orbital parameters with Equation (\ref{eqn:K}),
the same principle is valid for the transit light modulation, which can be modeled by specific functions presented in~\cite{2002ApJ...580L.171M}.
The shape of the light curve is dependent, among other things, on the inclination of the planetary orbital plane through the so-called ``impact parameter'', \textit{b} (i.e., the projected separation between the center of the stellar disk and the center of the planet disk), according to the relation $b=\frac{a \cos i}{R_{\star}}$, where \textit{a} is the semi-major axis of the orbit and $R_{\star}$ is the stellar radius (see, e.g., \cite{2010arXiv1001.2010W}). Once the inclination is measured, the real mass of the planet can be obtained. This is just an example of the synergy among the different planet detection techniques (e.g., \cite{2013pss3.book..489W}).}

The RV method is more sensitive to a large planetary mass and a short orbital period: actually, a~Jupiter-like planet orbiting at 1 AU from its host star produces an RV semi-amplitude of about 30~m~s$^{−1}$, while an Earth-like planet at the same distance induces an RV modulation of only
9 cm s$^{−1}$ (see also Table 1 in \cite{2010exop.book...27L} for an overview of the typical semi-amplitude values induced by different types of planets).
An RV survey also requires an intense observational effort: the observing baseline should be long enough to sample at least one, up to many periods of the planetary companion. 

Current state-of-the-art spectrographs in the visible (VIS) range provide an RV accuracy of about 1~m~s$^{-1}$ (see Section \ref{sec:vis}), allowing one in principle to detect small-mass exoplanets around solar-type stars with relatively short orbital periods. 
Anyway, the search for low-mass planets is hampered by the presence of the stellar activity that induces an intrinsic RV variation with similar amplitude with respect to the amplitude of the Keplerian signal, or even larger. 
 {Recently, M dwarfs have been identified as the most interesting objects to investigate, because of the higher planet occurrence rate in their habitable zones (Section \ref{nir}) and the more favorable ratio between planet and stellar masses. On the other hand, they are red stars, fainter in the VIS wavelength range.}
To overcome the limitations imposed by the stellar activity {and to allow a proper measurement of their spectra}, observations in the near-infrared (NIR) range have been proposed, since in this band, the RV modulations caused by intrinsic stellar phenomena are expected to be mitigated with respect to the VIS. For sure, observations of M dwarfs (or active stars, in general) with dedicated NIR spectrographs provide major advantages, but obtaining simultaneous VIS-NIR observations would allow reaching the next level, in terms of the completeness and ``speed'' of the information (Section \ref{sec:goal}).

Nowadays, astronomical research greatly benefits from multi-wavelength observations of the same phenomenon. In the field of exoplanets, particular attention to the NIR band has been payed in the last decade for the characterization of exoplanet atmospheres, both with the transit and direct imaging techniques (even in combination with the VIS range), while high-resolution high-precision spectroscopy in the NIR is still in its infancy, because of technical and technological issues that have been solved only recently. Today, thanks to to the efforts of coordinated teams in the creation of instruments like CARMENES or GIARPS
 (Section \ref{new}), it is possible to exploit combined VIS-NIR spectroscopic observations and start to investigate a parameter space of the exoplanet research previously not accessible.

In this review, I will draw an overview of the historical, technical and scientific scenarios that led to the current achievements in the framework of multi-band high-resolution spectroscopy in the field of exoplanet search. Those results represent the foundations of the future new-generation~facilities. 
 
\section{The Path toward Multi-Wavelength Observations} \label{first}
In this Section, I summarize the steps to reach high-precision RVs in the VIS range with high-resolution spectroscopy and how the expertise gained in this field helped to set up the first experiments in the NIR band, after the realization that this domain can give a significant contribution to the search for exoplanets. I shall also discuss the first combined VIS-NIR observations and what kind of benefits could be obtained with the VIS-NIR simultaneity.

\subsection{The VIS Regime} \label{sec:vis}
\begin{figure} [H]
 \begin{center}
 \begin{tabular}{c} 
 \includegraphics[height=5cm,angle=0]{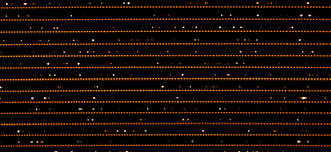}
	\end{tabular}
	\end{center}
	\vspace{-12pt}

 \caption[] { \label{fig:fp} HARPS echellogram showing the Fabry--Perot emission lines (dense and evenly spaced) and the features of the ThAr lamp. Reproduced with permission from \cite{2011SPIE.8151E..1FW}.}
 \end{figure}

After the historical discovery of 51\,Peg b, the search for exoplanets around FGK spectral type stars became one of the hottest topics in astrophysical research. That event pushed the scientific community to develop cutting-edge technologies to pursue a sufficient accuracy in the RV measurements, aiming to search for planets as similar as possible to the Earth. As a consequence, a new generation spectrographs was designed, allowing a good instrumental stability, and more accurate wavelength calibration techniques were studied. Since then, a number of ground-based surveys dedicated to the search for exoplanets with the RV method have been arranged, and up to the outstanding achievements of the transits survey performed by the NASA-{Kepler} satellite \cite{2010Sci...327..977B,2016RPPh...79c6901B}, the RV was the most productive method for exoplanets' detection (e.g., \cite{2011arXiv1109.2497M}). 
The technology developed for the optical instrumentation showed an increasing improvement through the years: the fiber-fed ELODIE (installed at the 1.93-m telescope of Observatoire de Haute-Provence -- OHP, France) spectrograph, which was used to find 51\,Peg b in 1995, already ensured a good RV accuracy at that time ($\sim$13 m s$^{-1}$), by using the simultaneous thorium reference technique \cite{1996A&AS..119..373B}. With this method, a scientific fiber feeds the instrument with the light of the target, while a second fiber is illuminated by a thorium-argon (ThAr) lamp, aiming to obtain a simultaneous wavelength calibration able to track and correct any instrumental drifts occurring during the observing night.
Other spectrographs, such as HIRES(High-Resolution Echelle Spectrometer, mounted at the 10-m Keck I telescope, USA), adopted the Iodine cell technique \cite{1996PASP..108..500B}, in which a gas absorption cell is inserted in the optical path of the instrument, obtaining a superimposition of the I$_2$ lines on the scientific spectra, producing thus a stable wavelength reference. This technique allows one to reach an RV precision up to 3 m s$^{-1}$ (e.g., \cite{2017AJ....153..208B}). 
A significant step forward in the accuracy of the RV measurement in the optical band has been achieved with HARPS (High Accuracy Radial velocity Planet Searcher) \cite{2000SPIE.4008..582P}, a high-resolution fiber-fed echelle spectrograph mounted at the ESO 3.6-m Telescope in La Silla (Chile). This spectrograph operates in a vacuum enclosure in controlled temperature and pressure environment, located in an isolated room to avoid vibrations connected to the telescope building. 
In order to obtain even more precise RV measurements, new calibration systems were studied and {routinely} adopted for HARPS and HARPS-N (its twin in the Northern Hemisphere; see Section \ref{giarps}), such as the Fabry--Perot (FP) interferometers \cite{2011SPIE.8151E..1FW}. As~for the ThAr method, a {calibration} source {(e.g., a lamp or laser)} overimposes well-spaced emission lines on the detector, close to the stellar spectrum. As an example, Figure \ref{fig:fp} displays a 2D spectrum of HARPS showing the traces of the FP compared to the less rich and unevenly-spaced lines of the ThAr lamp~\cite{2011SPIE.8151E..1FW}, demonstrating the high potential of this type of calibration.
A similar, but more sophisticated and expensive technique, the laser frequency comb has been developed and tested, as well \cite{2012SPIE.8446E..1WL,2014SPIE.9147E..1CP}. This method still needs a better optimization before its full exploitation, but {thanks to the excellent calibration performances, it should contribute to} providing an RV precision significantly lower~than~1~m~s$^{-1}$. 
 \vspace{-6pt}

Thanks to the state-of-the-art calibration techniques and to an optimized RV extraction \cite{2000SPIE.4008..582P}, HARPS has been able to reach an RV accuracy of 1 m s$^{-1}$ or less (depending on the SNR of the spectra and the stellar properties) and gave a significant contribution to exoplanet research in the last decade \cite{2011A&A...534A..58P,2013A&A...549A.109B,2014A&A...566A..35S}. 
HARPS and HARPS-N have represented the state-of-the-art of the RV technique for a long time: now, at the beginning of the scientific operations of ESPRESSO (Echelle SPectrograph for Rocky Exoplanet and Stable Spectroscopic Observations, \cite{2017arXiv171105250G}, \url{https://obswww.unige.ch/Instruments/espresso/}) at the Very Large Telescope (VLT, in Chile), the incredible regime of 10 cm s$^{-1}$ is waiting to be covered.
The {interested} reader is referred to \cite{2016PASP..128f6001F} for a detailed overview of the RV surveys of the last decades, as well as a resume of the status of the RV method.

With the previous discussion in mind, it should be emphasized that the planet distribution obtained by the RV surveys \cite{2008PASP..120..531C,2011arXiv1109.2497M,2016ApJ...819...28W} suffers from a sample selection bias.
Indeed, for years, the optical surveys have been mostly focused on ``quiet'' stars, i.e., stars with low or negligible stellar activity. Several processes can be ascribed to the stellar activity, such as the pulsations (timescale of a few tens of minutes for solar-type stars), granulation (hours) and magnetic activity (up to a few tens of days). All of them are responsible for the intrinsic RV variation of a star, {together with all the other sources of stellar noise with no periodic patterns, called ``jitter''}. In particular, magnetic activity induces a variable number of evolving photospheric inhomogeneities (e.g., cool spots and bright \textit{plage}), the visibility of which is modulated by the stellar rotation. These phenomena produce subtle time-dependent changes in the line profiles of the star, resulting in an RV modulation that can be erroneously interpreted as a Keplerian motion \cite{1997ApJ...485..319S,2005PASP..117..657W,2014A&A...566A..35S,Haywood2016}. For this reason, stellar activity represents a real problem in the search for exoplanets, and several studies have been proposed to identify useful diagnostics helping in discriminating the real nature of an RV variation.
First of all, a line deformation due to activity provokes an asymmetry of the line profile that can be observed through the line bisector, as shown, e.g., in \cite{2005A&A...442..775M}. In order to quantify the extent of the asymmetry, it is useful to evaluate the difference between average values of velocities in regions at the top and at the bottom parts of the line profile (see Figure 5 in \cite{2001A&A...379..279Q}), which is defined as the bisector velocity span (BVS~\cite{1988ApJ...334.1008T}). When photospheric inhomogeneities are responsible for the RV variation, the associated line profile variation produces a correlation between the RV and bisector span \cite{2001A&A...379..279Q}. Besides the bisector, other activity indicators have been proposed, aiming to find false-positives in the RV time series: some~of them take into account the asymmetry of the line profiles \cite{2013A&A...557A..93F,2018arXiv180407039L} or measure the emission of the H and K lines of the Ca II \cite{1984ApJ...279..763N,2010ApJ...725..875I}, centered at 3968.47 and 3933.66 \si{\angstrom}, respectively. 
Unfortunately, the lack of correlation does not imply the presence of a planet \cite{2017A&A...599A..90B}, and the correlation is not always linear or easily detectable \cite{2007A&A...473..983D}.
The treatment of the stellar activity can be very challenging, especially when one looks for low-mass planets having a small RV semi-amplitude comparable to or lower than the typical value of the {RV variation due to the stellar activity}. Only in recent years have specific tools been presented and heavily used to disentangle the planetary signal from the activity one; for instance, the Gaussian processes \cite{2014MNRAS.443.2517H}. This is the reason why several stellar classes have been generally not considered in the past for the planet searching surveys, with a consequent lack of information about the planet frequency around all those targets showing high levels of stellar activity (e.g., young and intermediate-age stars, M dwarfs, etc.). Today, the treatment of stellar activity requires a substantial number of data, an optimized observing strategy with intense monitoring and a good degree of precision of the measurements. {In principle, these conditions can be reached in the visible band (e.g., \cite{2016AJ....152..204L,2018arXiv180208320D}), as discussed above, while a less favorable situation takes place in the case of the NIR facilities.} 

\subsection{The Role of the NIR} \label{nir}
\begin{figure} [H]
 \centering
 \begin{tabular}{c} 
 \includegraphics[height=5.5cm,angle=0]{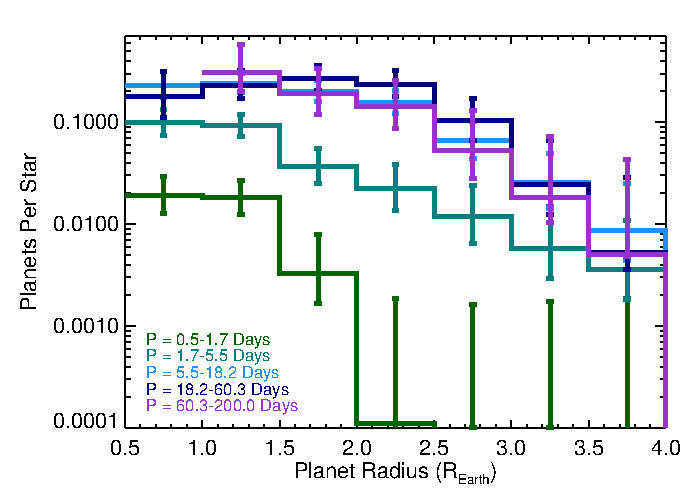}
 \includegraphics[height=5.5cm,angle=0]{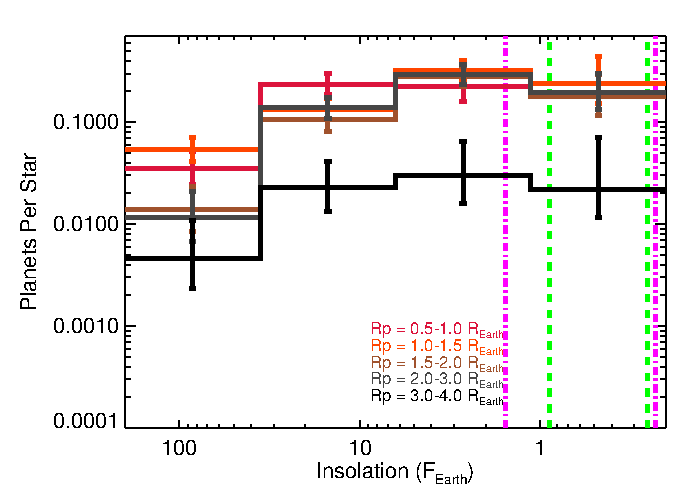}
	\end{tabular}
	\vspace{-12pt}
 \caption[] { \label{fig:dressing} Left panel: Occurrence rate of planets around M dwarf stars as a function of planet radius for all the {Kepler} candidates considered in \cite{2015ApJ...807...45D}. Right panel: Planet occurrence rate vs. insolation. The green box indicates the typical range of insolation of the habitable zone. The green dashed lines indicate the limits for the maximum greenhouse and moist greenhouse insolation, while the magenta dotted-dashed lines represent the less conservative recent Venus and early Mars limits. Credit \cite{2015ApJ...807...45D} \copyright AAS. Reproduced with permission.}
 \end{figure}

A drastic turnaround occurred when the planet occurrence rate was assessed for low-mass stars in the {Kepler} sample, according to which the formation of small planets is highly {favorable} with respect to FGK stars \cite{2012ApJS..201...15H,2013ApJ...767...95D}, {with an estimate of $2.5 \pm 0.2$ planets per M dwarfs with radii 1--4 R$_{\oplus}$ and periods shorter than 200 days \cite{2015ApJ...807...45D}}; see Figure \ref{fig:dressing}, left panel. Since the RV semi-amplitude \textit{K} (Equation (\ref{eqn:K})) scales inversely with the stellar mass and the planet orbital period, planets with similar properties induce a larger RV amplitude on an M dwarf with respect to more massive stars.
 \vspace{-6pt}
 
The lower luminosity of M dwarfs also implies that their habitable zone (HZ), i.e., the circumstellar region where water is allowed to exist in liquid state on a planet surface, lies at shorter orbital distances~\cite{2013ApJ...765..131K}. As a result, even planets with relatively short orbital periods could be located inside the HZ of an M dwarf \cite{2016PhR...663....1S}. Their occurrence rate has been estimated as {0.16$^{+0.17}_{−0.07}$ for Earth-sized planets (0.5--1.4 R$_{\oplus}$) and 0.12$^{+0.1}_{-0.05}$ for Super Earths
(1.4--4 R$_{\oplus}$) in the case of the moist greenhouse inner and maximum greenhouse outer limits; 0.24$^{+0.18}_{−0.08}$ for Earth-sized planets and 0.21$^{+0.11}_{-0.06}$ for Super Earths for the recent Venus and early Mars limits, as defined in \cite{2013ApJ...765..131K}}; see the right panel of Figure \ref{fig:dressing}.
Thanks to their characteristics and their ubiquity in the Universe \cite{2015AJ....149....5W}, M dwarfs appeared as the ideal shortcut to find the first habitable Earth-like planet. 
As a downside, M dwarfs are extremely faint in the VIS wavelength and {show increasing levels of stellar activity from earlier to later spectral types}. This is where the NIR wavelength band comes to the rescue \cite{2010ApJ...710..432R}: the M dwarfs' spectral energy distribution peaks between 1 and 2.5 $\upmu$m, and the observation in the NIR range provides important advantages. 

The main benefit when observing in the NIR band is that the stellar RV {variation due to activity} is expected to be lower than in the VIS \cite{2010ApJ...710..432R,2015ApJ...798...63M}, because of the lower intensity contrast between the stellar surface and the spots \cite{1989A&A...211...99B,2011MNRAS.412.1599B}. In the framework of the search for exoplanets, this represents an outstanding advantage: 
while the activity signal is wavelength dependent, the RV modulation due to a planetary companion must show the same amplitude regardless of the band. As a matter of fact, this allows discriminating between signals of different origin. 

The search for habitable rocky planets around M dwarfs is not the only science case that benefits from observation in the NIR. 
The extension of the high-precision radial velocity measurements to this spectral range allows one to search for planetary companions to all those objects that were usually not considered due to the intrinsic difficulties, like stars with a moderate or high level of stellar activity, intermediate-age stars in open clusters or stars in very young associations (a few Myrs). In the latter case, exoplanets could be observed when their formation was ongoing, and the determination of their orbital parameters allows one to put constraints on the planet migration mechanisms \cite{2008ApJ...686..580C,2014prpl.conf..667B,2016ApJ...829..114B}. 
According~to the recent results \cite{2016Natur.534..662D,2017MNRAS.467.1342Y} the frequency of planets around young stars is higher with respect to the older ones. A systematic investigation of these objects and the verification of this finding could help to understand the evolution of the systems. Comparison between the planetary mass distribution also allows inferring the photo-evaporation of the planet's atmosphere \cite{2016ApJ...820L...8M,2016A&A...589A..75M}. 
With the support of the NIR band, it is also possible to search for the presence of exoplanets around red giant stars.
The~typical pulsation periods of red giants range between two regimes: a short-time periodicity driven by pressure modes (with a timescale of the order of a few hours up to days, e.g., \cite{1994ApJ...432..763H}) and a long-time periodicity (hundreds of days) probably due to oscillatory convective modes \cite{2015MNRAS.452.3863S} (see \cite{2013ARA&A..51..353C} for a description of stellar oscillations in red giants). Obviously, this could hamper the detection of planets with similar orbital periods \cite{2018AJ....155..120H}, but the oscillations being chromatic phenomena \cite{2001PASP..113..983P}, it is possible to distinguish them with combined VIS-NIR observations.

Finally, alongside the RV search for exoplanets, high-resolution NIR spectroscopy also allows the atmospheric characterization of close-in giant planets, representing nowadays the training ground for the future detection of biosignatures of exo-Earths \cite{2016frap.confE..61C}. Actually, the main molecular constituents of a planet atmosphere show their absorption bands in the near- and mid-infrared (water, oxygen, methane, carbon monoxide, etc.; see, e.g., \cite{2013Icar..226.1654T}) and can be detected through the transmission and emission spectroscopy method. In the former, the light of the host star is filtered by the atmosphere of a giant gas planet during the transit phase, impressing its extremely weak features on the stellar spectrum \cite{2010Natur.465.1049S,2013MNRAS.432.1980R,2018arXiv180109569B}. In the latter, the thermal emission from the planet, coming from internal heat and/or the reflected starlight, can be extrapolated from the stellar spectra by following part of the orbital phase, even for non-transiting exoplanets \cite{2012ApJ...753L..25R,2015A&A...576A.134M,2017AJ....154...78P}. This review is focused on the planet search with the RV method, so the atmospheric characterization will be marginally discussed; anyway, the {interested} reader can refer to \cite{2010ARA&A..48..631S,2015ApJ...804...10C} for a detailed introduction to this argument.

It is now clear that the science achievable with the extension of the high-resolution spectroscopy to the NIR domain is really noteworthy. Unfortunately, all the previous efforts to develop techniques and build high-precision instruments working efficiently in the optical range could not be easily applied in another band.
In the NIR range, the thermal noise represents an annoying source of {disturbance} in the spectra, especially toward the redder parts. For this reason, the instruments must be cooled down, which implies different technology and more costs. However, the main challenge to obtain high-precision RV measurements in the NIR is to define a reliable procedure for an accurate wavelength calibration. The simultaneous ThAr technique, so efficient in the VIS, is not suitable because in the NIR range, thorium shows a small number of faint spectral lines, which are generally contaminated by the scattered light of the argon emission lines, sensitive to changes in the environmental conditions (see \cite{2009ApJ...692.1590M} and the references therein). 
The calibration with a laser comb or FP is studied for NIR, as well. Both the hardware and a corresponding optimized technique were not available up to a few years ago, but after the promising results of dedicated studies \cite{2008Natur.452..610L,2008Sci...321.1335S}, more efforts were performed and will be/are adopted in some of the current and future NIR facilities \cite{tozzi2016,2017Msngr.169...21B} (see also Sections \ref{giarps} and \ref{nirps}). 

\begin{figure} [H]
 \begin{center}
 \begin{tabular}{c} 
 \includegraphics[height=6cm,angle=0]{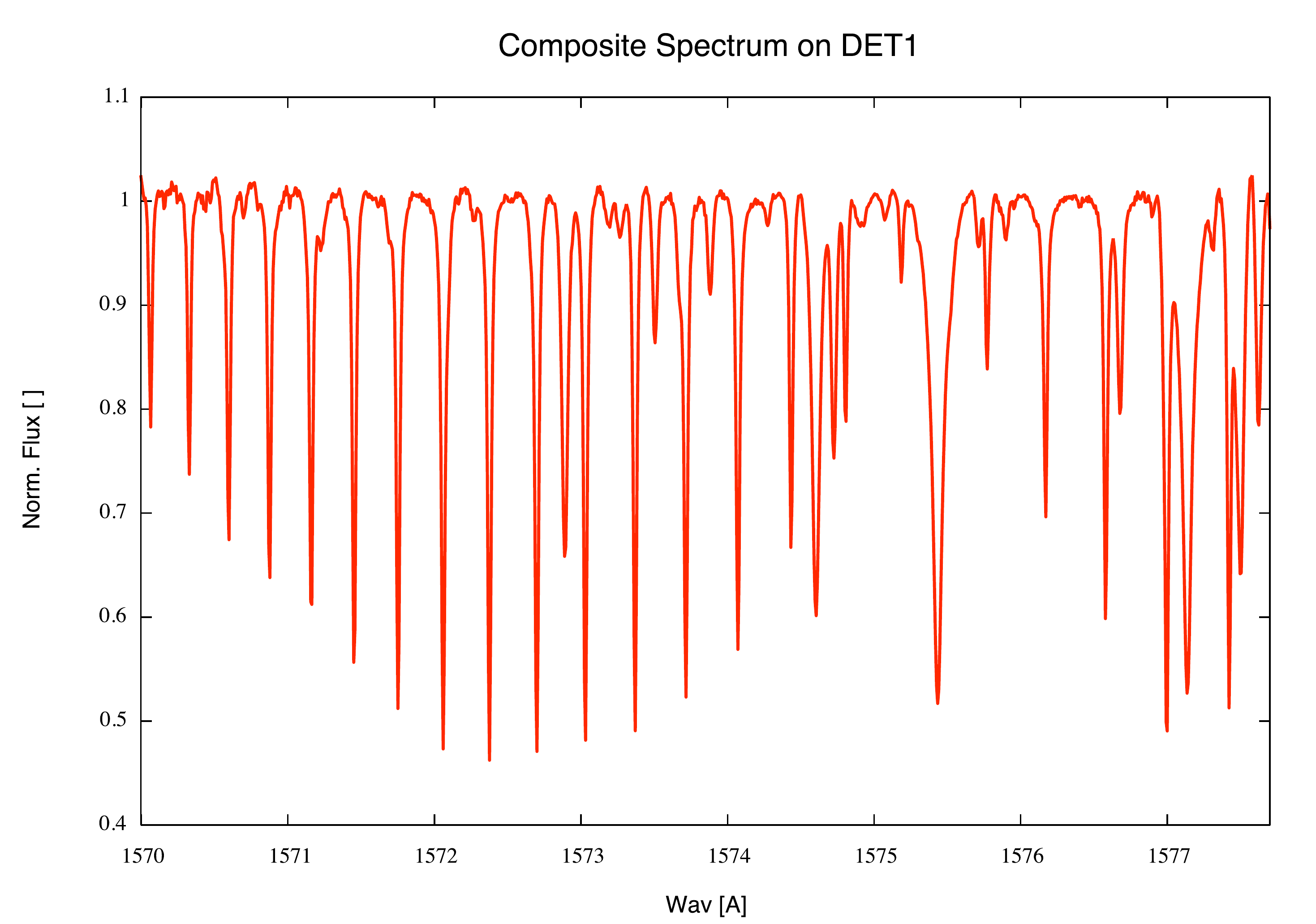}
	\end{tabular}
	\end{center}
	\vspace{-12pt}
	
 \caption[] { \label{fig:tell} Portion of the stellar spectrum of the RV standard HD\,108309 obtained with the detector 1 (DET1) of the NIR spectrograph CRIRES (see Sect. \ref{specnir}). Telluric features due to the absorption of molecular compounds of the Earth's atmosphere dominate the wavelength range between the J and H bands. Credit: \cite{2010A&A...511A..55F}, reproduced with permission \copyright ESO.}
 \end{figure}

Many studies have been carried out about the optimization of the NIR gas absorption cell technique \cite{2008A&A...491..929S,2009ApJ...692.1590M,2010ApJ...713..410B,2012PASP..124..586A}, similar to the I$_2$ cell in the visible. This approach has been demonstrated to be quite effective (RV precision, from $\sim$20 m s$^{-1}$ \cite{2008Sci...321.1335S} up to 3--5 m s$^{-1}$ \cite{2010ApJ...713..410B}), even if the emission lines of a selected gas are usually available for specific wavelength bands only (e.g., J or H) and do not cover the whole spectral range typically offered by a spectrograph. In the last few years, dedicated studies have been performed aiming to define suitable gas mixtures to be inserted in the cells, able to span larger wavelength ranges \cite{2014SPIE.9147E..5GS}.
Finally, another approach to obtain a reliable wavelength reference for RV measurement is to exploit the rich forest of telluric lines that dominate the NIR spectra \cite{1973MNRAS.162..255G} (see, e.g., Figure \ref{fig:tell}). These lines are due to the light absorption of molecular species of the Earth's atmosphere (mainly OH, H$_2$O, CO$_2$) that impress their features on the spectra just like an absorption gas cell. This technique produced satisfying results, with RV precision up to 5--10 m/s \cite{2010A&A...511A..55F}. The drawback of this method is that the telluric lines are dependent on atmospheric conditions such as temperature, humidity and wind \cite{amt-6-2893-2013}. On the other hand, the presence of the dense telluric features reduces the number of spectral orders useful for the RV measurement. Therefore, several methods have been proposed {to} correct for the effect of those lines from the stellar spectra, allowing one to work with a cleaner and optimized spectrum (e.g.,\cite{2014A&A...561A...2A,2014SPIE.9149E..05A,2018PASP..130g4502S}). 

\vspace{-18pt}

\subsubsection*{Spectrographs for the NIR Spectral Domain} \label{specnir}
\begin{figure} [H]
 \begin{center}
 \begin{tabular}{c} 
 \includegraphics[height=17cm,angle=0]{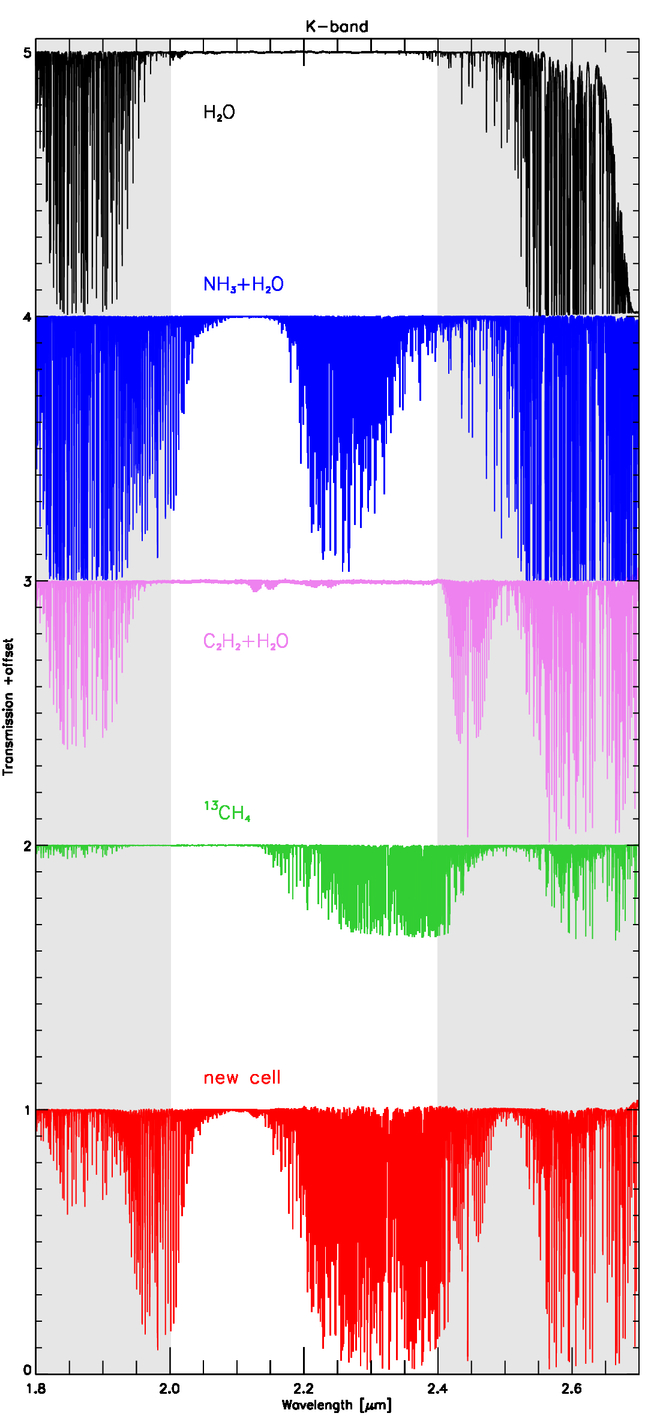}
	\end{tabular}
	\end{center}
	\vspace{-12pt}
 \caption[] { \label{fig:cell} Absorption spectrum of the CRIRES+ cell (in red), optimized in the K-band, as the result of a suitable mixture of absorption gas species (black, blue, pink and green lines). Reproduced with permission from \cite{2014SPIE.9147E..5GS}.}
 \end{figure}

Starting in the 1990s, the first examples of high-resolution echelle spectrographs in the NIR band became available, allowing the pioneering studies of, e.g., young stars, M and L dwarfs and red giants (Section \ref{pioneer}). However, the RV uncertainties were of the order of hundreds of m s$^{-1}$, preventing the detection of possible low-mass companions. 
Among those instruments, the Cryogenic Echelle Spectrograph (CSHELL) \cite{1993SPIE.1946..313G} at the NASA 3-m Infrared Telescope Facility
(IRTF, {Maunakea Observatory, USA}) covered a wavelength range between 0.95 and 5.4 $\upmu$m, allowing low-, intermediate and high-resolution modes (up to 40,000). 
The same spectral coverage is also offered by the Near InfraRed SPECtrograph (NIRSPEC) \cite{1998SPIE.3354..566M} at the Keck II 10-m telescope, providing a high-resolution of 25,000 (a low-resolution mode of R $\sim$2200 is also available). 
The Phoenix spectrograph \cite{2003SPIE.4834..353H} was previously mounted at the telescopes of the Kitt Peak Observatory before it was adapted for the 8 m Gemini South Telescope ({Cerro Pachon, Chile}). Its resolving power ranges between 50,000 and 80,000, and it covers the NIR range between 1 and 5 $\upmu$m. 
For years, the state-of-the-art of NIR spectrographs were represented by CRIRES (CRyogenic high-resolution InfraRed Echelle Spectrograph) \cite{2004SPIE.5492.1218K} at the ESO 8-m at VLT, operational since April 2007. CRIRES provided R = 100,000 in the wavelength range 0.95--5.38 $\upmu$m. However, {since it was a single-order spectrograph}, the single exposure was limited to a narrow spectral range ($\sim$1/70 of the central wavelength), resulting in a demanding observing request to obtain a sufficient wavelength coverage for the measure of reliable RVs. Nevertheless, several authors obtained remarkable results in the field of exoplanets thanks to very good RV precision by using different techniques \cite{2010ApJ...713..410B,2010A&A...511A..55F}. To~fulfill the increasing request of RV search for planets observing programs, a few years ago, a significant refurbishment has been proposed for CRIRES in order to improve the observing efficiency, giving~rise to the CRIRES+ project \cite{2014SPIE.9147E..19F}. The introduction of a {full} cross-dispersing element, as well as larger detectors will ensure an extension of the wavelength coverage for a single spectrum ($\sim$1/7 of the central wavelength), by maintaining both the previous spectral coverage and the high resolution. 
Moreover, the new equipment foresees the insertion of an absorption gas cell filled with a suitable mixture of gas molecular species (acetylene, ammonia and an isotopologue of the methane) able to cover both the H and K bands. Since the region of maximum spectral emission of M dwarfs is located in the K band, the CRIRES+ cell is particularly optimized in that range (Figure \ref{fig:cell}). According to the simulations, the use of the cell would allow reaching an RV measurement precision of $\sim$3 m s$^{-1}$ \cite{2014SPIE.9147E..5GS}.
 
 \vspace{-18pt}

\subsection{First Multi-Wavelength Investigations} \label{pioneer}
All the spectrographs previously mentioned operate in the VIS ``or'' in the NIR bands. Since the second half of the 2000s, some authors proposed the exploitation of the combined VIS-NIR information to obtain a more comprehensive description of the stellar activity. Actually, the science cases exploiting the NIR high-resolution spectroscopy illustrated in Section \ref{nir} obtain an even greater advance when studied with both VIS and NIR spectral bands.
Those studies show the actual potential of this method, even if the first NIR spectrographs were not optimized for high-precision RV measurement. 

\begin{figure} [H]
 \begin{center}
 \begin{tabular}{c} 
 \includegraphics[height=9cm,angle=0]{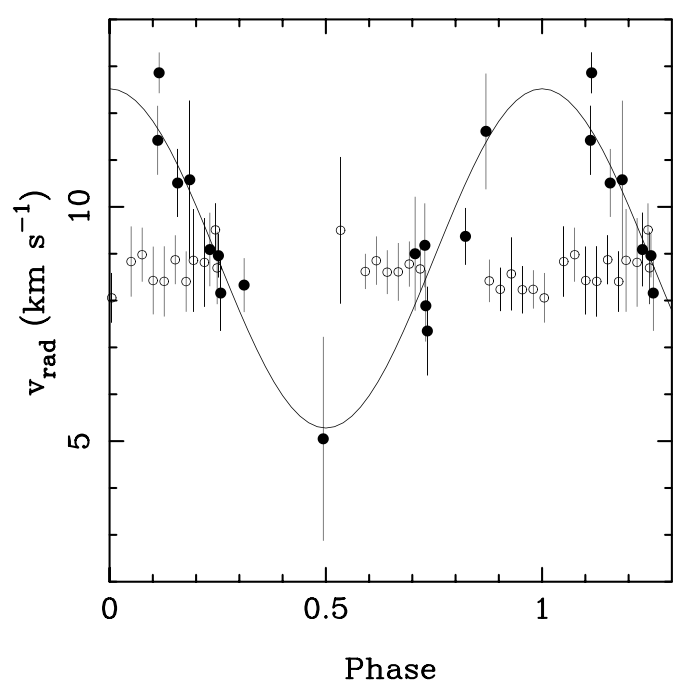}
	\end{tabular}
	\end{center}
	\vspace{-12pt}
 \caption[] { \label{fig:martin} Comparison between RV measurements in the visible (black circles) and NIR RV (open circles) of the brown dwarf LP 944-20. The black line is the model of the clear modulation shown by VIS data. Credit: \cite{2006ApJ...644L..75M} \copyright AAS. Reproduced with permission.}
 \end{figure}

The first investigation \cite{2006ApJ...644L..75M} that compared VIS and NIR RV datasets involved the observation of the brown dwarf LP 944-20. VIS data, obtained with UVES (UV-Visual Echelle Spectrograph)  at VLT  showed the significant RV dispersion of $\sim$3.5 km s$^{-1}$ and a periodic pattern every 3.7 h, suggesting the presence of a giant planet companion very close to its host (Figure \ref{fig:martin}, black circles). Anyway, both the RV amplitude and the period were unambiguously ruled-out by NIR data from NIRSPEC: their flat behavior indicates that the observed variation is not due to a reflex motion, but it is rather related to the fast rotation period of the brown dwarf. The presented NIR RV dispersion is $\sim$360 m s$^{-1}$, in full agreement with the prediction that activity-related processes are wavelength dependent and less prominent in the NIR with respect to the VIS band.
On the other hand, the observations presented by \cite{2006ApJ...644L..75M} were not simultaneous, so~they could not attest to the actual situation of the stellar activity at the time of the measurements, but they are surely conclusive about the fact that the signal observed in the VIS is not consistent with a giant planet orbiting the target. As Figure \ref{fig:martin} shows, the uncertainties on the single RV are quite large ($\geq$500~m~s$^{-1}$ for both the spectrographs), since the two instruments are not optimized for high-precision RV measurements. Anyway, due to the large amplitude of that specific signal, these uncertainties are appropriate to perform the proposed analysis.
 \vspace{-6pt}
 
A similar approach was applied to a small sample of three young ($\sim$2 Myrs) T Tauri stars \cite{2008ApJ...687L.103P} by using the VIS Coud\'e echelle spectrograph at the 2.7-m telescope at the McDonald Observatory and the CSHELL NIR spectrograph at the NASA IRTF 3-m telescope. One star of the sample shows a significant correlation between RV and bisector span, suggesting that the VIS RV modulation is due to the typical prominent activity of a young star, while the remaining targets show no correlation between RV and bisector.
As in \cite{2006ApJ...644L..75M}, the NIR data excluded planetary companions around the three targets, since the comparison between multi-wavelength RVs shows the same behavior as in Figure~\ref{fig:martin}. This~work demonstrates that the lack of correlation between RV and bisector span in this kind of objects is not sufficient to claim the presence of a planet, validating the simulation performed by~\cite{2007A&A...473..983D} that demonstrates the dependence of such a correlation from the stellar rotation velocity and the resolution of the spectrograph used.
Similar studies on young stars followed, including dedicated projects to search planets in their early stage of formation and evolution \cite{2012ApJ...761..164C} or showing that in the particular case of very large and long-lived cool spots, the RV amplitude observed in the K-band can be significantly larger than in the VIS \cite{2011ApJ...736..123M,2017ApJ...836..200G}.

Among those studies, the one that became the most emblematic example of the power of the multi-wavelength observations method was the retreat of the hot Jupiter claimed around the classical T Tauri star TW Hya (about 10 Myrs old). In this case, the giant planet was claimed with data from the visible spectrograph FEROS (Fibre-fed Optical Echelle Spectrograph, at the 2.2-m Max-Planck-Gesellschaft ESO telescope in La Silla, Chile), which~suggested the presence of such a companion with a period of 3.56 days and a mass of about 10~M$_{\rm Jup}$~\cite{2008Natur.451...38S}. Even in this case, the suspect RV variation was found to be related to the stellar activity by NIR observations with CRIRES \cite{2008A&A...489L...9H}.

As previously mentioned, combined VIS-NIR observations are also suitable to investigate the presence of exoplanets around intermediate-mass stars, in order to understand how stellar evolution affects planetary systems. The optical RV survey on K giants presented by \cite{2013A&A...555A..87M} allowed selecting low-mass planetary/brown dwarfs candidates, confirmed after NIR observations with CRIRES.

\subsection{The Goal of the Simultaneity} \label{sec:goal}
The studies presented in the previous section showed how the combination between VIS and NIR data provides useful information on the origin of RV signals. All of them were necessarily performed with data obtained from observations collected at different observatories, where the VIS and NIR instruments are available, possibly with a time separation of months or even years. This~means that a single research team should submit at least two different proposals to obtain telescope time with VIS and NIR spectrographs. Most important is the consequent loss of observing efficiency, since~the VIS-NIR dataset of the same target must be collected twice. from the scientific point of view, the~impossibility to obtain simultaneous VIS-NIR data prevents a correct interpretation of the behavior of stellar activity, since the two datasets monitor two different periods of the stellar activity cycle.
The~huge potential of the combined VIS-NIR observations has been caught in the last few years by several national and international consortia that started to design new facilities allowing simultaneous VIS-NIR high-resolution spectroscopy aiming to answer the new key questions of exoplanet research. CARMENES, GIARPS and NIRPS + HARPS, described in the next sections, are the first examples of the technological efforts made to fulfill the new scientific requirements.

\begin{figure} [H]
 \begin{center}
 \begin{tabular}{c} 
 \includegraphics[height=5.5cm,angle=0]{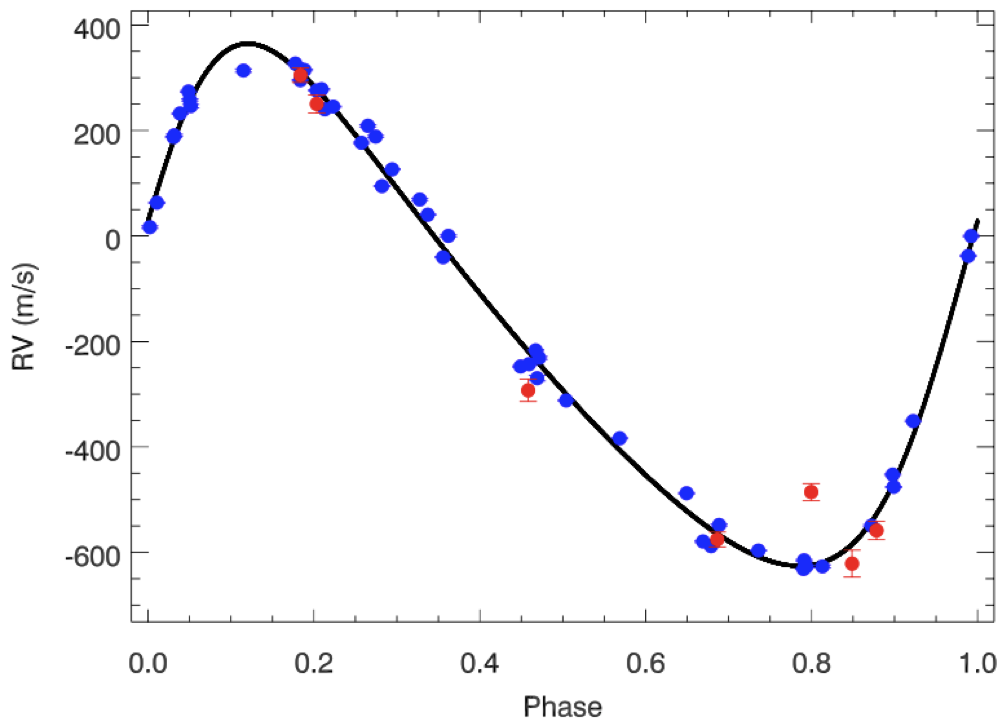}
 \qquad
 \includegraphics[height=5.5cm,angle=0]{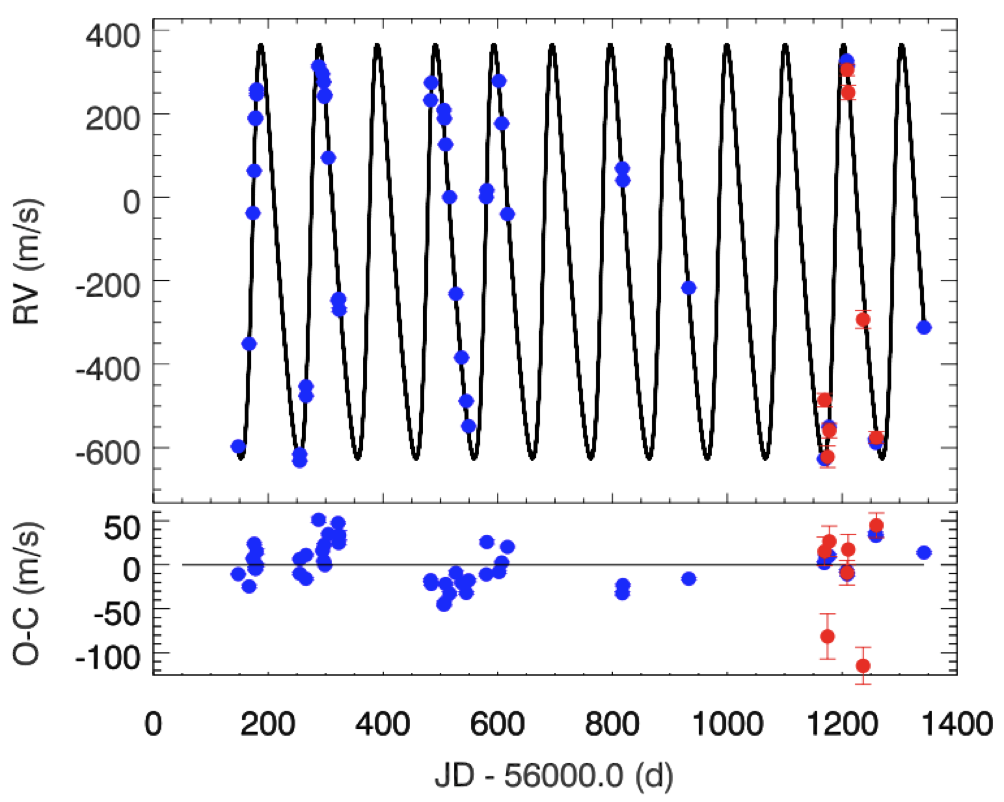} 
	\end{tabular}
	\end{center}
	\vspace{-12pt}
 \caption[] { \label{fig:kg7} Left panel: VIS (blue dots) and NIR (red dots) RV measurements of the giant star TYC 4282-605-1, phase-folded with the orbital period detected by \cite{2017A&A...606A..51G}. Right upper panel: Time series of the same RVs showing the quasi-simultaneity and agreement between the measurements in the two bands. Right lower panel: RV residuals between the observed points and the model. Credit: \cite{2017A&A...606A..51G} reproduced with permission \copyright ESO.}
 \end{figure}

The first result obtained from partially-simultaneous VIS-NIR observations regards the detection of a substellar companion with minimum mass $m_p\sin i$ = 10.78 $\pm$ 0.12 $M_J$ and an orbital period $P~=~101.54~\pm 0.05$ days around the K giant star TYC 4282-605-1 \cite{2017A&A...606A..51G}. This target was at first monitored with the VIS spectrograph HARPS-N at TNG, showing an RV modulation that could have been caused by a planet or stellar oscillations. A second dedicated monitoring has been arranged combining observations with HARPS-N and the NIR high-resolution spectrograph of TNG, GIANO, in its first configuration (see Section \ref{giarps}), obtaining thus quasi-simultaneous observations. Since the two spectrographs were mounted at the two opposite Nasmyth foci of the telescope, a certain amount of time was necessary to change the telescope configuration, so the time difference between the VIS and the NIR spectra in the same night was about a few hours, which is a very short time compared to the period of the candidate. The two datasets showed that the VIS and NIR RV semi-amplitudes were compatible (Figure \ref{fig:kg7}), providing a ``quick and easy'' detection of the planet, avoiding performing two separate observing runs with a VIS and an NIR instrument.
This work is the very first example of cooperation between two spectrographs that would have been part of the GIARPS instrument, introduced in Section \ref{giarps}.
 \vspace{-18pt}
 
\section{New Breakthrough Facilities} \label{new}
A comprehensive overview of the current and future facilities providing high-precision RV in the visible and near-infrared bands (or both) is presented in \cite{2017RNAAS...1...51W} and partially resumed in Appendix \ref{sec:app} of this manuscript.
In the following, I will be focused on those instruments that allow simultaneous multi-band observations. Two of them are already operational, i.e., CARMENES (Section \ref{carmenes}) and GIARPS (Section \ref{giarps}), while others are ongoing, like HARPS + NIRPS (Section \ref{nirps}). All of them represent the groundwork of future facilities exploiting the large collecting area of new generation ELTs (Extremely Large Telescopes; Section \ref{future}).

\subsection{CARMENES at the 3.5-m Telescope at Calar Alto} \label{carmenes}
The primacy of the high-resolution spectrograph CARMENES (Calar Alto high-Resolution search for M dwarfs with exo-Earths with Near-infrared and optical Echelle Spectrographs) \cite{2014SPIE.9147E..1FQ} is to be the first instrument to be specifically designed to provide simultaneous VIS-NIR observations. Moreover, the main science case of this spectrograph, i.e., the detection of low-mass rocky planets in the habitable zone of M-dwarfs, is openly declared in its name.
The final concept of CARMENES had been established a few years after the first proposal in 2005. {The first study considered only an NIR spectrograph, called NAHUAL (Near-infrAred High-resolUtion spectrogrAph for pLanet hunting) \cite{2005AN....326.1015M}, presented as one of the instruments for the 10.4-m GTC (Gran Telescopio de Canarias) in La Palma (Canary Islands). After several modifications and the merging with a similar spectrograph and science driver, but in the VIS range, the CARMENES Project, equally~led by a Spanish -- German collaboration ({the list of all the Institutions belonging to the CARMENES Consortium is available here: \url{http://carmenes.caha.es/ext/consortium/index.html}}), was proposed and selected in 2009 for the 3.5-m telescope of the CAHA Observatory ({Centro Astron\'omico Hispano-Alem\'an de Calar Alto, Almeria -- Spain}).
A more detailed overview of the CARMENES history is provided in \cite{2013hsa7.conf..842A}.
The commissioning of CARMENES occurred at the end of 2015, lasting for six weeks \cite{2017hsa9.conf..599A}.

The instrument is composed of two similar highly-stabilized, high-resolution echelle spectrographs, with an optical and optomechanical design presented in \cite{2012SPIE.8446E..33S}. The visible channel covers the wavelength range from 0.55--0.95 $\upmu$m with a spectral resolution R $\sim$94,000, while~the NIR channel covers the 0.95--1.7 $\upmu$m interval with R $\sim$80,000.
The thermal stability is guaranteed by a dedicated cooling system ensuring constant temperature inside vacuum tanks in which the two spectrographs are located, within few hundredths of a degree in 24 h \cite{2016SPIE.9912E..62B}. While the VIS instrument operates at room temperature, the NIR one requires a cooler environment of about 140 K. 
The light coming from the target is split with a dichroic, which feeds the VIS and NIR channels through fibers. Besides the scientific fiber, a second one is used to feed the spectrographs with a simultaneous calibration unit, by using an emission-line lamp or an FP (or the sky background in the case of faint objects).

Thanks to the monitored and stabilized environment and the simultaneous calibration technique, CARMENES is able to reach an RV precision of 1 m s$^{-1}$, necessary to reach the objectives of its primary science, as verified in the commissioning run \cite{2017hsa9.conf..599A}. Figure \ref{fig:carm_all} shows the full simultaneous VIS and NIR spectrum of an M-dwarf obtained during the CARMENES commissioning. This plot demonstrates the huge information content of such a kind of instrument. The data reduction is assigned to the pipeline CARACAL (CARMENES Reduction And Calibration) included in the CARMENES data flow \cite{2016SPIE.9910E..0EC}, while the RVs from CARMENES spectra are obtained with the dedicated code, SERVAL (SpEctrum Radial Velocity Analyser) \cite{2018A&A...609A..12Z}, which creates a template spectrum for each star, starting from a number of observed spectra, and performs an iterative least-squares fitting between the template and the single spectrum. 
After the commissioning, in 2016, the Guaranteed Time Observations (GTO) program, which has been assigned 750 nights over five years, has started its operations to monitor a sample of more than 300 M dwarfs \cite{2018A&A...612A..49R} (see Section \ref{sec:mdwarf} for a summary of the first results).
Since the main goal of CARMENES is the detection of low-mass rocky planets in the HZ of M dwarfs, a wide recognition of literature data and specific analyses have been necessary for a careful selection of the sample. 
The coordinated effort for the characterization of the input catalog included: ({i}) the analysis of low resolution spectra to derive spectral types, surface gravity, metallicity and chromospheric activity for all the stars \cite{2015A&A...577A.128A}; ({ii}) the census of binary companions to the stars through high-resolution images \cite{2017A&A...597A..47C}; ({iii}) the measurement of stellar rotation velocity and stellar activity indices \cite{2018arXiv180202102J}.
The complete database, named CARMENCITA (the CARMENES Cool dwarf Information and daTa Archive), is composed by $\sim$2200 M dwarfs \cite{2015A&A...577A.128A}, of which $\sim$300 were discarded due to stellar multiplicity. Finally, the GTO sample includes 324 stars \cite{2018A&A...612A..49R} from the CARMENCITA catalog. 

 \begin{figure} [H]
 \begin{center}
 \begin{tabular}{c} 
 \includegraphics[height=7cm,angle=0]{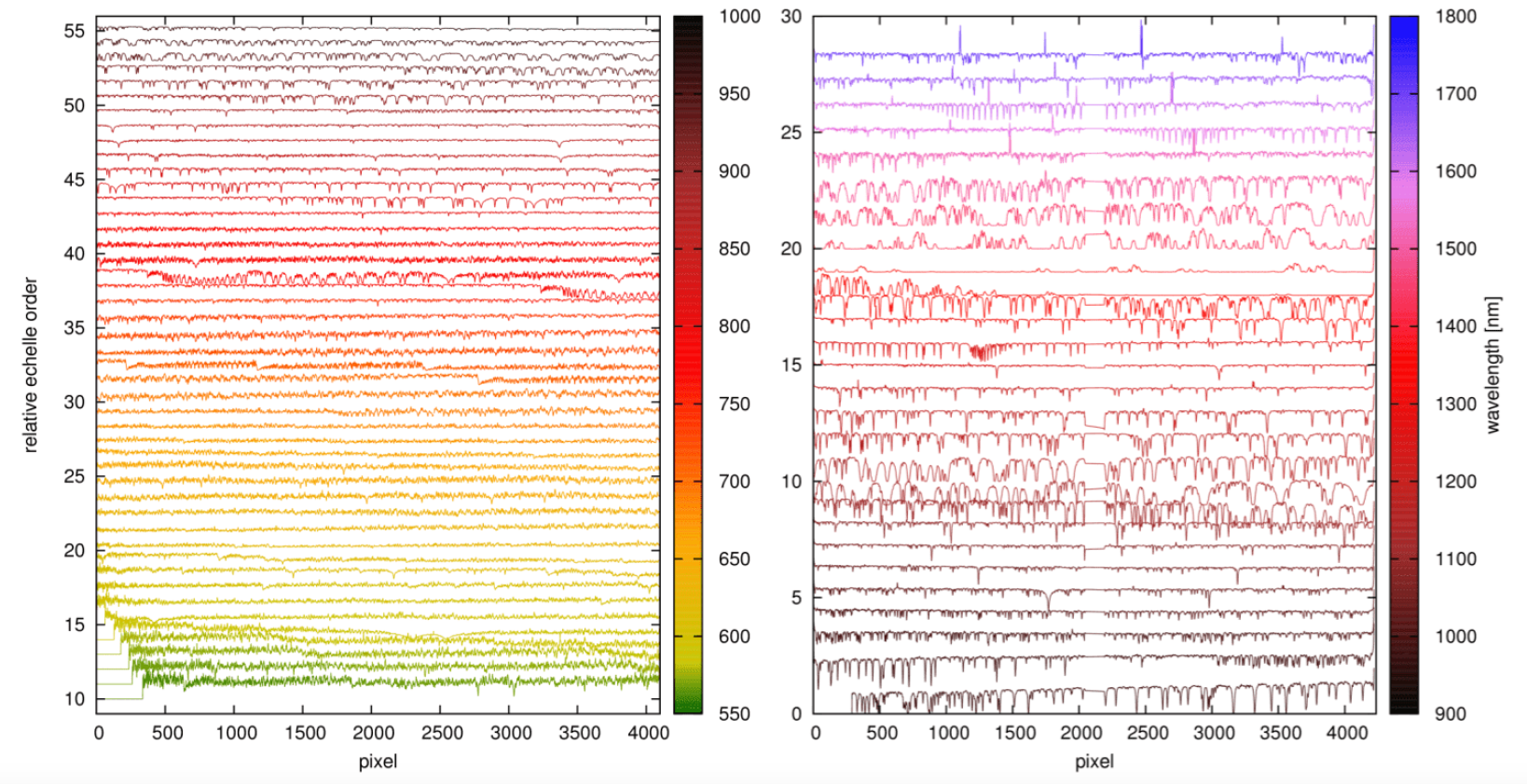} 
	\end{tabular}
	\end{center}
	\vspace{-12pt}
 \caption[] { \label{fig:carm_all} The complete VIS and NIR spectrum of an M dwarf star provided by CARMENES. Reproduced with permission from \cite{2017hsa9.conf..599A}.}
 \end{figure} 

\vspace{-18pt}

\subsection{GIARPS at TNG} \label{giarps}
In 2012, an agreement between the Italian National Institute for Astrophysics (INAF) and the HARPS-N Consortium (see \url{https://plone.unige.ch/HARPS-N/overview}) established the installation of the HARPS-N \cite{cosentino2014} spectrograph at the 3.6-m INAF-Telescopio Nazionale Galileo (TNG) in La Palma. HARPS-N is a cross-dispersed echelle spectrograph working in the visible range, between $0.39$ and $0.68$ $\upmu$m, with a spectral resolution R = 115,000. It is mounted at the Nasmyth-B focus of the TNG and operates in an extremely stabilized and monitored environment, allowing one to reach a typical radial velocity precision of 0.8 m\,s$^{-1}$ \cite{2016PASP..128f6001F}. 
At the same time, a second high-resolution echelle spectrograph, GIANO \cite{oliva2006}, has been mounted at TNG, as well. It works in the NIR range, from 0.95--2.45 $\upmu$m, at a spectral resolution of 50,000. Despite it being designed to be mounted at the Nasmyth-B focus receiving a direct light feed from the telescope, it was finally installed in the other focus chamber (Nasmyth-A) due to the scheduling constraints of TNG. In that configuration, the cryogenic vessel containing GIANO was detached from the telescope, which implied the use of ZBLAN fibers to feed the spectrograph, maintaining the wide wavelength range of the instrument, from the Y--K band. As a drawback, the adopted fiber introduced a modal noise \cite{2015MmSAI..86..490I} in the acquired spectra, which significantly degraded the overall instrumental efficiency \cite{origlia2014}. 

A few years later, the coordinators of the Progetto Premiale WOW, ``A Way to Other Worlds'' ({a funding scheme of the Italian Ministry of Education, University and Research promoting the cooperation and the connection of the Italian extrasolar planets community through the realization of several projects.}), proposed a full exploitation of the potential of HARPS-N and GIANO, aiming to obtain a facility providing simultaneous VIS-NIR high-resolution spectroscopy.
To reach this goal, it~was necessary to join the two spectrographs by moving GIANO and allowing it to share the same focus of HARPS-N. A feasibility study to allow the direct feeding of GIANO from the telescope has been performed and included the definition of a new pre-slit system \cite{tozzi2016} and the building of a platform attached to the telescope fork to host the cryogenic dewar. Thanks to the new setup and the consequent removal of the fibers, the instrument, now called ``GIANO-B'', shows the expected efficiency \cite{claudi2018}. 
After this operation, GIANO-B and HARPS-N are able to work together in the GIARPS (GIAno-b and haRPS-n, see Figure \ref{fig:giarps:bd}) \cite{2017EPJP..132..364C} configuration thanks to the insertion of a dichroic in the light path that splits the entrance beam into the VIS and NIR components. An entrance slider allows choosing the preferred observing mode: HARPS-N only, GIANO-B only and GIARPS. 

A uranium-neon lamp is used to obtain the wavelength calibration of the GIANO-B spectra, but the optimization of an FP interferometer is ongoing to provide simultaneous calibration and better accuracy. A preliminary study to build a gas absorption cell for GIANO-B has also been performed~\cite{ulf}, with similar characteristics with respect to the gas cell for CRIRES+.
The data reduction is performed in real time by the pipeline GOFIO \cite{rainer2018}, which processes calibration images and scientific spectra during the observing night. 
 \begin{figure} [H]
 \begin{center}
 \begin{tabular}{c} 
 \includegraphics[height=6cm,angle=0]{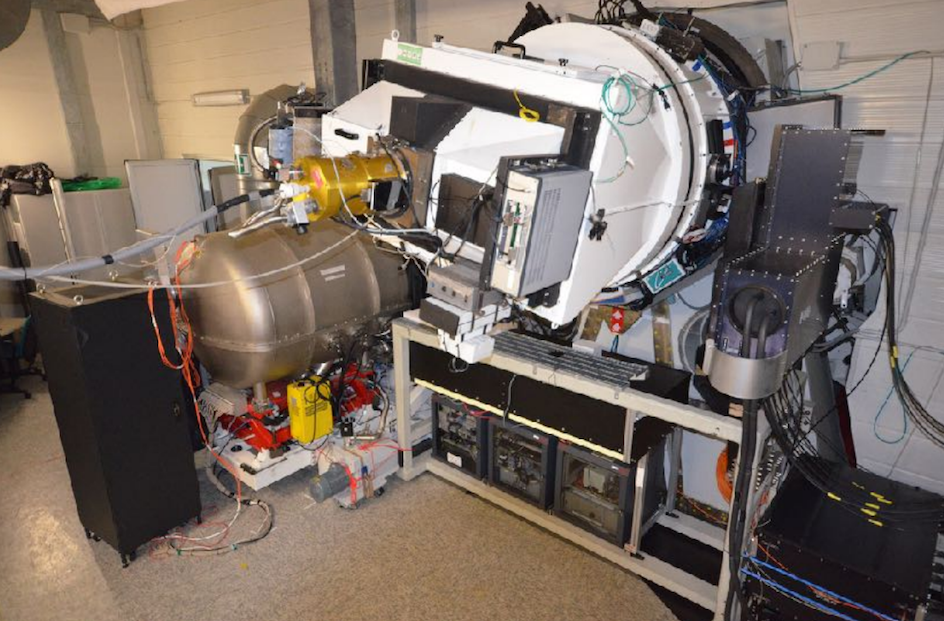}
	\end{tabular}
	\end{center}
	\vspace{-6pt}
 \caption[example] 
 { \label{fig:giarps:bd} 
 The GIARPS configuration at the Nasmyth-B focus of TNG. {Left}: The electronics and the dewar containing GIANO-B: the latter is mounted on a dedicated platform attached to the telescope fork. {Bottom center}: The new GIANO-B preslit (black box) and the corresponding electronics. {Right}:~The~front-end and calibration units of HARPS-N. GIANO-B and HARPS-N receive the NIR and VIS light beams separated by a dichroic, hosted by LRS (Low Resolution Spectrograph, {top center}). Credit:~TNG-Fundaci\'on Galileo Galilei.}
 \end{figure} 
The commissioning of GIARPS occurred in three separate runs between August 2016 and March 2017, and the new facility has been offered by TNG since April 2017. 

Due to the versatility of an instrument such as GIARPS, several science cases have been identified~\cite{2017EPJP..132..364C}, but the main scientific driver is the search and characterization of exoplanets.
Together~with the TNG and the Observatories of Arcetri and Padova ({the  full list of institutes belonging to the GIARPS Consortium are listed on the web page: \url{http://www.tng.iac.es/news/2017/04/04/giarps/}}), the~Italian community GAPS (Global Architecture of Planetary Systems; see, e.g., \cite{2016frap.confE..69B}) had a significant role in the building of GIARPS, as well as in its scientific validation. In order to exploit the capabilities of the instrument, an observing large program has been proposed by GAPS, aiming to study the frequency of hot young planets around stars having a young (age < 0.1 Gyrs) and intermediate age (<0.7 Gyrs) and to characterize hot Jupiters'/Neptunes' atmospheres. First observations are ongoing in parallel with the instrument characterization.
In this framework, the instrumental stability of the red arm of GIARPS (i.e., GIANO-B) has been tested, since the one of HARPS-N is already well known and robust. To this aim, systematic observations of an RV standard have been collected over one semester in GIARPS mode. HD\,3765 is a bright quiet star (K2V, V mag = 7.36, H mag = 5.27) with a low {intrinsic RV variation} of $\sim$2.4 m s$^{-1}$ \cite{2010ApJ...725..875I}. {The NIR RVs are obtained as the weighted average RV for each exposure and its corresponding uncertainty, producing an internal error of 23~m~s$^{-1}$ (see~also \cite{2016ExA....41..351C} for a detailed description of the RV extraction from GIANO spectra and Section 3.1 in~\cite{2018A&A...613A..50C} for an update)}, while the VIS counterpart shows internal errors of 0.3 m s$^{-1}$. {This gap must be mainly ascribed to the different instrumental stability and different wavelength calibration methods, which are extremely optimized for HARPS-N.} Figure \ref{fig:HD3765} shows the resulting RVs, both in the VIS (blue open diamonds) and NIR (red dots) ranges. While the VIS RVs present an rms scatter of 2.3~m~s$^{-1}$, in agreement with previous observations, {the NIR RVs present an rms scatter of 17 m s$^{-1}$ over a semester, representing the preliminary long-term stability of GIANO-B. It must be taken into account that the external contributions to HARPS-N RV noise, e.g., guiding errors, photon noise, wavelength calibration, are generally lower or much lower than 1 m s$^{-1}$, while these effects have more impact on the GIANO-B stability. In this case, some temperature drift can occur, and the wavelength calibration, obtained from the U-Ne lamp, cannot provide the same RV precision as for HARPS-N. The forthcoming use of an FP should help to limit this source of noise.}
\vspace{-12pt}

 \begin{figure} [H]
 \begin{center}
 \begin{tabular}{c} 
 \includegraphics[height=8cm]{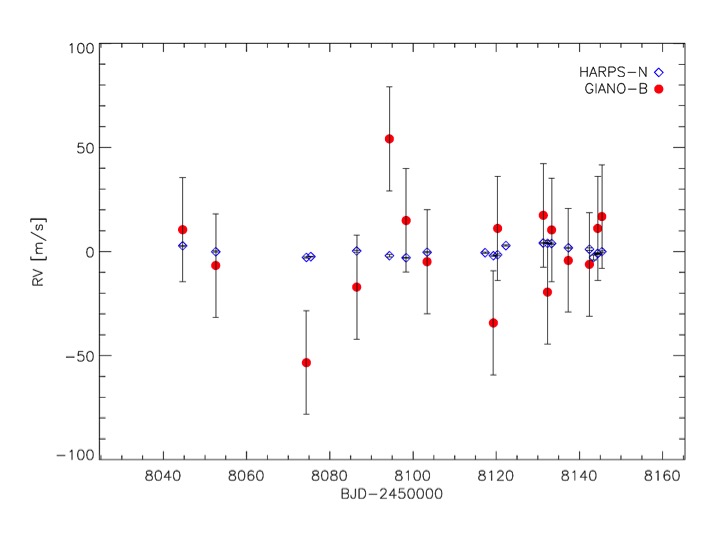}
 	\end{tabular}
 	\end{center}
 	\vspace{-18pt}
 \caption[example] 
 { \label{fig:HD3765} 
Radial velocities of the standard star HD\,3765 from simultaneous GIANO-B (red points) and HARPS-N (blue open diamonds) spectra in GIARPS configuration. Courtesy of I. Carleo. }
 \end{figure} 

\vspace{-18pt}

\subsection{Work in Progress: NIRPS + HARPS at ESO} \label{nirps}
\textls[-15]{NIRPS (Near Infra Red Planet Searcher) \cite{2017SPIE10400E..18W} is going to be the NIR counterpart of the HARPS spectrograph at the ESO 3.6-m telescope. The leadership of the building consortium is shared by the Geneva University (to exploit the expertise gained in the building of HARPS) and Montr\'eal~University ({the full list of the NIRPS consortium can be found at~\url{http://www.eso.org/public/teles-instr/lasilla/36/nirps/}}). 
Similarly to CARMENES, NIRPS was initially proposed for a different telescope, as an answer to the 3.58-m ESO NTT (New Technology Telescope in La Silla, Chile) call for new instruments in 2014. At that time, another project was selected (the NIR medium-resolution spectrograph SOXS (Son Of X-Shooter ~\cite{2016SPIE.9908E..41S}), but considering the high scientific impact of an instrument such as NIRPS, in particular, if coupled with HARPS, the ESO Board asked the proposers to adapt the instrumental design from the Ritchey--Chretien configuration of the NTT to the Cassegrain of 3.6 m. 
The wavelength coverage of NIRPS ranges between 0.98 and 1.80 $\upmu$m (from the Y--H band), renouncing thus to the K band to favor the simplicity of the design, even if a possible extension to that range is not excluded in the future.}

Just like HARPS, NIRPS will be installed inside a cryostat (Figure \ref{fig:nirps}) with controlled temperature and pressure, ensuring an accurate monitoring of the instrumental drift. 
NIRPS will benefit from a dedicated adaptive optic (AO) system \cite{2016SPIE.9909E..41C} that allows reducing the light beam size and, as a direct consequence, the size of the instrument due to a more simple optical design. The nominal configuration foresees the concentration of the starlight into a fiber corresponding to 0.4 arcsec on the sky, producing a spectral resolution R = 100,000. In the case of faint targets or bad viewing condition, it will be possible to use a fiber with a larger field of view, 0.9, for a more efficient photon collection, with R = 75,000. This is the first practical high-performance attempt to combine AO with a fiber-fed high-resolution spectrograph.
 {The cooperation between the dedicated AO system (that will scramble the modes at the fiber entrance) and the combination of three scrambling techniques (octagonal fibers, fiber stretcher and a double scrambler) allows mitigating} the typical modal noise \cite{2017SPIE10400E..18W} that affected, e.g., GIANO in its previous configuration (Section \ref{giarps}).
Users could choose between a laser frequency comb and a stabilized FP calibration system to obtain very high-precision RV. Indeed, the expected RV accuracy of NIRPS will not be very different from HARPS, reaching 1 m s$^{-1}$ in less than 30 min for an M3 star with an H magnitude of 9 \cite{2017Msngr.169...21B}. The overall wavelength coverage of the NIRPS + HARPS configuration will be 0.4--1.8 $\upmu$m, except for a small window between 0.7 and 0.95 $\upmu$m {(i.e., the gap between the two instruments)} that is used for the wave-front sensing of the AO system. NIRPS's first light is expected by~2019.
 \begin{figure} [H]
 \begin{center}
 \begin{tabular}{c} 
 \includegraphics[height=6cm,angle=0]{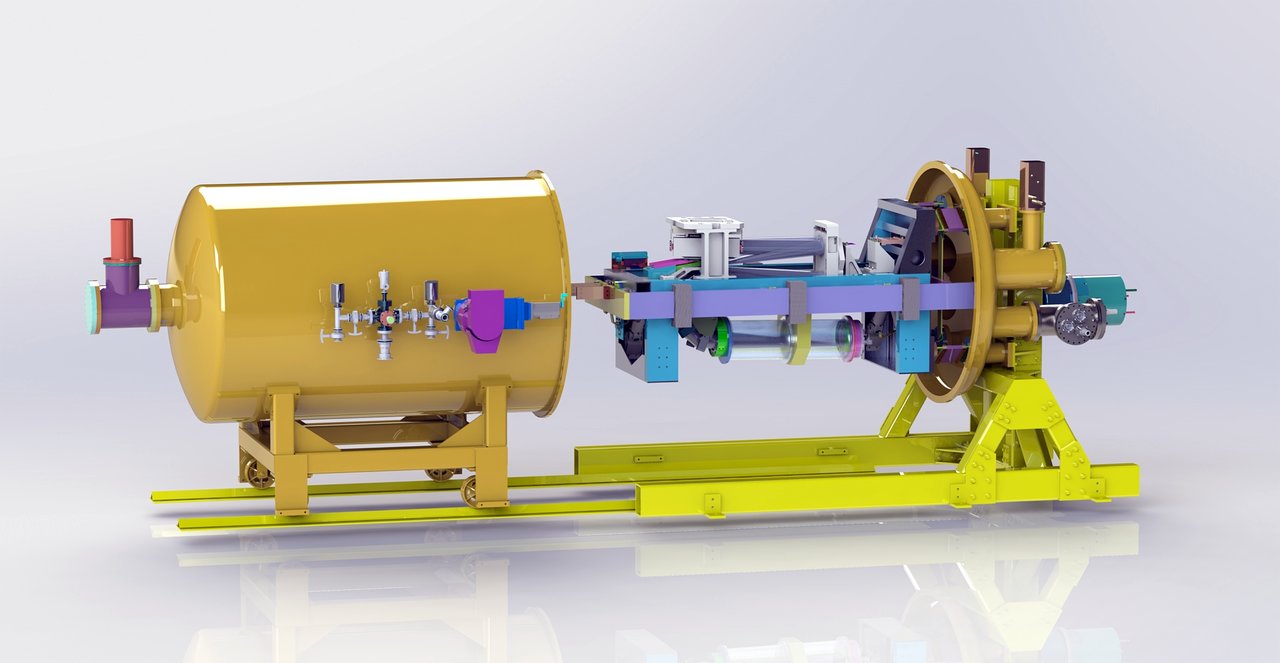}
	\end{tabular}
	\end{center}
	\vspace{-12pt}
 \caption[example] 
 { \label{fig:nirps} Cryostat and optical bench design for the NIRPS spectrograph. Credit: ESO website \url{https://www.eso.org/public/images/ann17056a/}.}
 \end{figure} 

The three science cases identified for the five years of GTO of NIRPS (combined with HARPS) are: ({i}) the characterization of transiting Earth-mass planets around M dwarfs; ({ii}) the selection of candidates for high-contrast imaging performed by ELT class telescopes; ({iii}) atmospheric characterization of exoplanets.
The NIRPS + HARPS combination is then going to play an outstanding role in the context of the full characterization of potentially habitable planets in the near future. As soon as the first M dwarf planet candidates from the NASA TESS satellite (Transiting Exoplanet Survey Satellite, see \url{https://tess.gsfc.nasa.gov/science.html} \cite{2014SPIE.9143E..20R}) are released, an RV follow-up with NIRPS + HARPS would provide their masses. Since the planet radius is measured with the transit observation, it is then possible to place those values in the mass-radius diagram (see, e.g., Figure 4 in \cite{2015ApJ...800..135D}), which is the first step to reveal the planet composition, even if a very small uncertainty is required to avoid degeneracy among different models. Those planets could subsequently be subject for atmospheric characterization with JWST (James Webb Space Telescope, {\url{https://jwst.stsci.edu/science/science-corner/white-papers}}), and their host stars could be investigated to search for further companions on wider orbits through imaging observations or through the astrometry method by using the huge Gaia database (\url{http://sci.esa.int/gaia/}).

Finally, investigations of molecular features of hot Jupiters and hot Neptunes in the combined wavelength domains will be performed. 
As in the case of the other mentioned facilities, these studies will allow us to create a sort of chemical species database, useful for future investigations, as well as to validate laboratory experiments \cite{2016MmSAI..87..104C}. Anyway, the most important aim of such a type of detection is to explore and/or refine the techniques to recover atmospheric planetary signals waiting to investigate the atmosphere of potentially habitable rocky planets by using high-resolution VIS-NIR spectrographs installed at telescopes with larger collecting area.

\subsection{Future Instrumentation} \label{future}
The exploitation of the huge collecting area of the future Extremely Large Telescopes (ELTs) represents an unprecedented opportunity and feeds great expectations for the search for exoplanets and characterization science cases. The quest for an Earth-analog planet around a solar-like star is the primary objective of the new-generation instruments, and the synergy among the next big international projects is going to make the difference. 
Among those instruments, the Giant Magellan Telescope (GMT, built by a consortium including American, Australian and Korean institutions, see \url{https://www.gmto.org/}) \cite{2014SPIE.9145E..1CB} is a 25.4-mdiameter segmented mirror telescope that will be based at Las Campanas Observatory (Chile), while the European ELT (E-ELT, an ESO facility, see {\url{https://www.eso.org/sci/facilities/eelt/}}) is a 39-m diameter segmented mirror telescope, which will be located at Cerro Armazones (Chile). Both of them are expected to be commissioned in 2024 and are planned to mount VIS and NIR high-resolution spectrographs.

\subsubsection{G-CLEF and GMTNIRS at the GMT} \label{gmt}

The Giant Magellan Telescope is going to be equipped by a first-light VIS and NIR high-resolution echelle spectrographs: G-CLEF (GMT Consortium Large Earth Finder, see \url{https://www.gmto.org/resources/visible-echelle-spectrograph-g-clef/})  and GMTNIRS (GMT Near-IR Spectrograph, {\url{https://www.gmto.org/resources/ir-echelle-spectrograph-gmtnirs/}}).

\begin{itemize}
\item G-CLEF is a fiber-fed, cross-dispersed echelle spectrograph built for general purposes, but designed to provide also high precision RVs (up to 10 cm s$^{-1}$ \cite{2014SPIE.9147E..26S}). 
The wavelength range covered by G-CLEF spans between 0.35 and 1 $\upmu$m to satisfy the requirements of the main science cases. The bluer part is included to perform studies on very metal-poor stars since the iron content is evaluated through very faint features between 3500 and 3900 \si{\angstrom}, undetectable from the existing facilities in terms of high resolution and telescope aperture. The wavelength coverage is pushed up to 1 $\upmu$m to exploit the spectral information of the M dwarfs and obtain precise RVs for these stars. 
Among the other exoplanet science cases, it is noteworthy to mention the search for biomarkers (in particular, the features of molecular oxygen between 7600 and 7700 \si{\angstrom} in the atmospheres of the exoplanets) and the follow-up of the planet candidates detected by the TESS~satellite.\\
The spectral resolution can be chosen according to the science program. G-CLEF offers both high-resolution modes (R $\sim$105,000), available with or without an image scrambler that enhances the RV precision and medium resolution modes (35,000 and 19,000).
As for GMTNIRS, particular attention is paid to providing the maximum performance in terms of throughput, to exploit the benefit of the huge telescope collecting area. To optimize the incoming stellar flux, the light beam will be split through a dichroic in two channels, red and blue (see Figure \ref{fig:gmtnirs}, upper panel), each~one equipped with specific detectors and coatings.
 \begin{figure} [H]
 \begin{center}
 \begin{tabular}{c} 
 \includegraphics[height=4.8cm,angle=0]{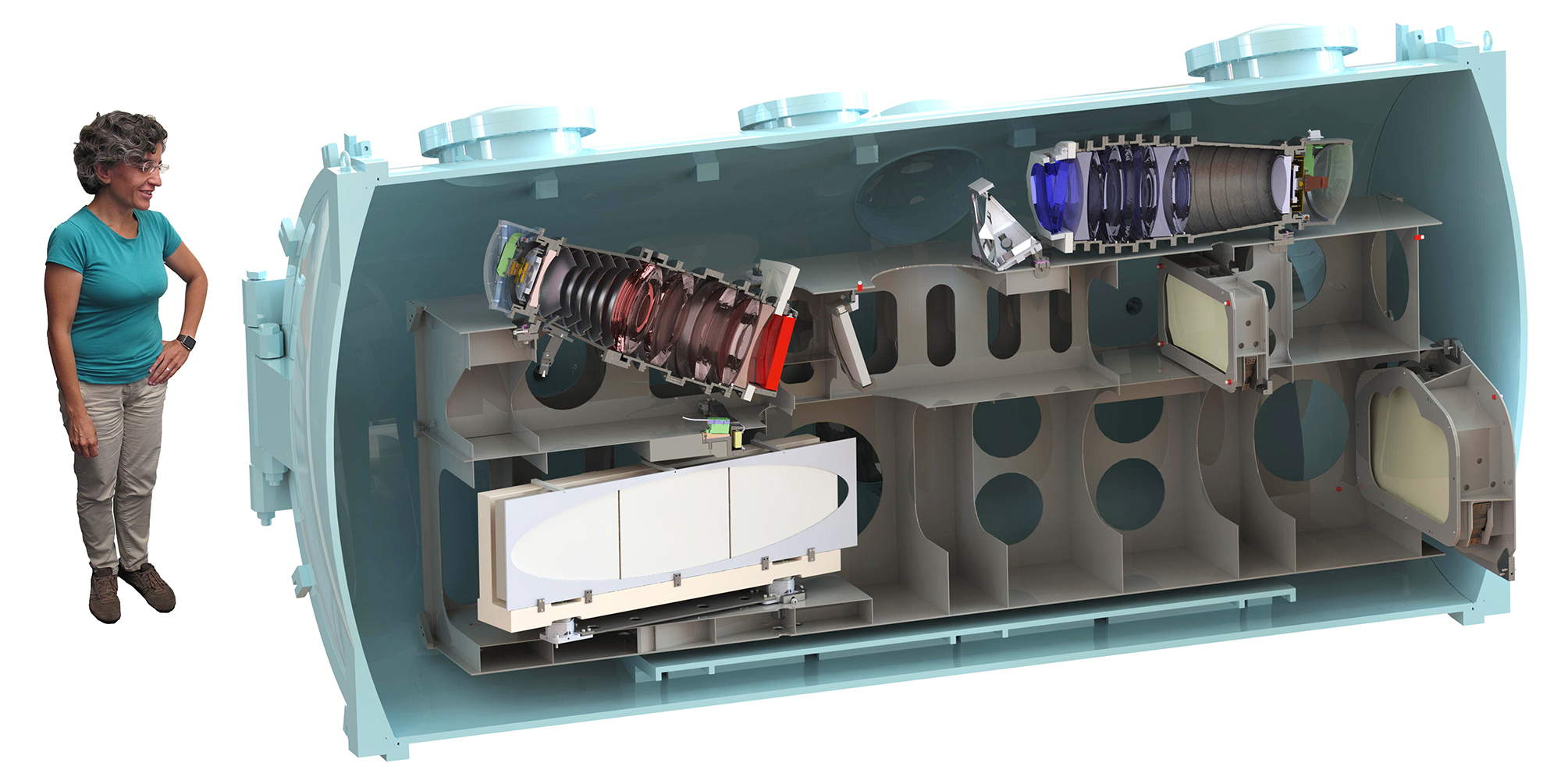}\\
 \includegraphics[height=5.3cm,angle=0]{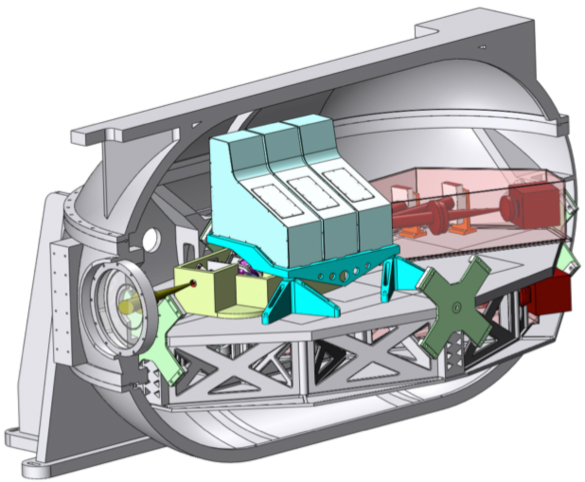}
	\end{tabular}
	\end{center}
	\vspace{-12pt}
 \caption[example] 
 { \label{fig:gmtnirs} Upper panel: Mechanical design of the G-CLEF spectrograph inside the vacuum vessel with the location of the blue and red cameras. Reproduced with permission of Prof. A. Szentgyorgyi and Dr. M. L\'opez-Morales. Lower panel: Same picture, but for the GMTNIRS cryostat. The spectrographs in the J, H~and~K bands are indicated as three light-blue boxes above the optical bench. 
 The L spectrograph is located on the top of the bench (red box), while the M band spectrograph is right under the L one, hidden here. Reproduced with permission of Dr. C. Brooks. Credits: The University of Texas at Austin and the Korea Astronomy and Space Science Institute.}
 \end{figure} 
 \vspace{-6pt}
\item GMTNIRS will cover the wavelength range between 1.15 and 5.3 $\upmu$m making use of five spectrograph units (see Figure \ref{fig:gmtnirs}, lower panel), each one dedicated to a specific atmospheric window: J, H, K, L and M bands. The light that feeds the instrument will be subdivided into the five channels thanks to a series of dichroic elements. The spectral resolution is expected to be $\sim$60,000 in the JHK bands and $\sim$85,000 in the L and M bands. 
In order to test the innovative technical elements planned to equip GMTNIRS, the University of Texas and the Korea Astronomy and Space Science Institute (KASI) joined to build a sort of forerunner instrument, the cross-dispersed NIR spectrograph IGRINS (Immersion Grating Infrared Spectrometer, see {\url{http://www.as.utexas.edu/mcdonald/facilities/2.7m/igrins.html}}) \cite{2016SPIE.9908E..0CM}. IGRINS, which has operated since 2014, has been installed alternatively at the Harlan J. Smith 2.7-m telescope (McDonald Observatory, USA), at the 4.3-m Discovery Channel Telescope (Lowell Observatory, USA) and at the 8.1-m Gemini South Telescope. It covers the H and K windows, from 1.45--2.5 $\upmu$m in a single acquisition with a resolving power of R = 45,000. The design of this instrument was particularly focused to optimize the throughput rather than to reach extreme precision in RV, and~the same concept will be adopted for GMTNIRS, as well. \\
The instrumental design originally proposed for GMTNIRS \cite{2010SPIE.7735E..2KL} is currently evolving \cite{2014SPIE.9147E..22J,2016SPIE.9908E..21J}, aiming to match the requirements of the main scientific driver emerging through the years, i.e., the observation of exoplanetary atmospheres, as for HIRES (see the discussion in Section \ref{hires}). Being part of a large and international project as the GMT, the GMTNIRS spectrograph will not be focused on exoplanet science alone; it will provide a significant contribution to the study of young stellar objects, debris disks, protoplanetary systems, stellar evolution, interstellar medium and the star formation history of the Galaxy.

\end{itemize}

These two instruments alone show the peculiarity of having a larger spectral coverage with respect to ``traditional'' VIS or NIR spectrographs. For this reason, a sort of multi-wavelength comparison can be performed even with data from a single instrument; anyway, their simultaneous use will be extremely interesting in the framework of exoplanet study.

\subsubsection{HIRES at E-ELT} \label{hires}
The main scientific driver of HIRES, the High Resolution Spectrograph \cite{2016SPIE.9908E..23M} for the E-ELT, is the atmospheric characterization of Neptune-like down to Earth-like planets that will be revealed by PLATO ({PLAnetary Transits and Oscillations of stars, an ESA M-class space mission dedicated to the transit search of thousands of planets, including the terrestrial ones, and asteroseismic study of the host stars, with an expected launch date in 2026}) \cite{2014ExA....38..249R}. Just like CARMENES, GIARPS and HARPS + NIRPS, the well-known potential of the combined VIS-NIR observations additionally coupled to a 40-m class telescope represents a great opportunity for exoplanets' characterization. The required accuracy for the RV precision is 10 cm s$^{-1}$.
HIRES is the result of the combination between the two AO-assisted high-resolution spectrographs proposed for the Phase A study of E-ELT: CODEX (COsmic Dynamics and EXo-earth experiment)
 \cite{2010SPIE.7735E..2FP}, operating in the VIS band (R = 120,000), and SIMPLE \cite{2010SPIE.7735E..2BO}, operating in the NIR (R = 130,000). The~new project thus foresees a unique instrument working in the wavelength range between 0.37 and 2.5~$\upmu$m with a resolving power of 100,000. 

Actually, as Figure \ref{fig:hires} shows, HIRES should be composed by different modules, allowing the subdivision of the costs according to the resources.
According to the preliminary design, there will be four fiber-fed cross-dispersed spectrograph modules, each one dedicated to a specific wavelength coverage: U-B, V-R-I, Y-J-H and K bands (the latter two would be cryogenic).
HIRES will surely be the most demanding effort in the building of high-resolution spectrographs, and as Table 2 in \cite{2016SPIE.9908E..23M} shows, it will be accomplished only with the collection of all the expertise gained in the production of the high-performance spectrographs realized in the last few years. {HIRES will be a second-generation instrument, so it will not be available at the E-ELT first light.}
The HIRES Consortium includes institutes from several countries (see {\url{https://www.arcetri.astro.it/~hires/consortium.html}}), and INAF is the leading technical institute.
The key science of HIRES will be the full investigation of planetary atmospheres (chemical composition, stratification and weather) of Neptune-like to Earth-like planets, including~those in the HZ of their host stars \cite{2013arXiv1310.3163M}. The detection of biosignatures is also expected. Since~a typical RV survey is very demanding from the point of view of time request, the use of E-ELT is not justified, since it will be available to a huge community having heterogeneous lists of relevant science cases. {The high pressure on E-ELT instrumentation should relax the time request of other facilities. Therefore, the RV monitoring for the search of exoplanets or characterization should proceed, e.g., with ESPRESSO or other future available spectrographs at 10-m class telescopes (see Appendix \ref{sec:app})}.
Anyway, the extreme RV precision ensured by HIRES could be exploited in the measurement of the Rossiter--McLaughlin (RM) effect, an RV anomaly in the RV curve detectable during the transit of a planet. This effect, which represents an independent confirmation of the actual presence of the planet, allows deriving the obliquity of the planetary orbit with respect to the stellar spin axis, which is an important observable to understand the planet migration history and/or constrain the star-planet tidal interactions \cite{2012ApJ...757...18A}.
 \begin{figure} [H]
 \begin{center}
 \begin{tabular}{c} 
 \includegraphics[height=7cm,angle=0]{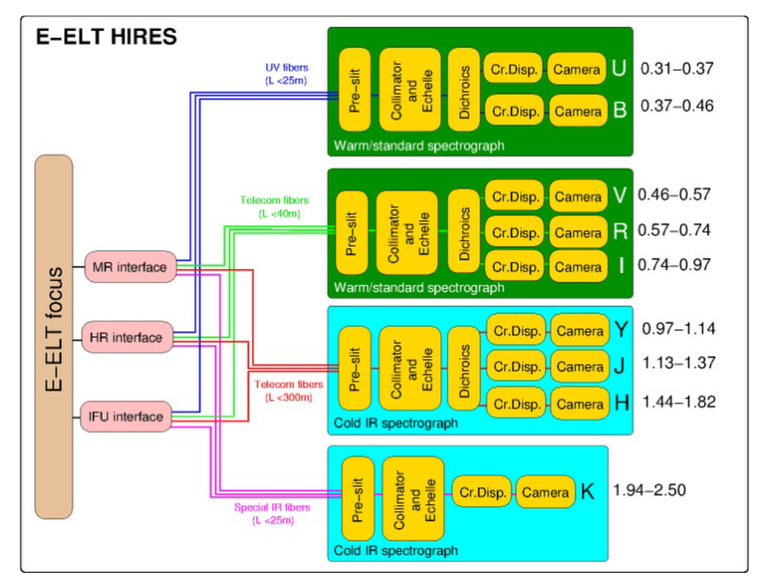}
	\end{tabular}
	\end{center}
	\vspace{-12pt}
 \caption[example] 
 { \label{fig:hires} Technical scheme summarizing the different modules of the HIRES spectrograph. Reproduced with permission from \cite{2016SPIE.9908E..23M}.}
 \end{figure} 
\vspace{-18pt}

\section{First Results} \label{results}
As discussed in Section \ref{pioneer}, combined VIS-NIR observations have already been proposed by several authors, who exploited the multi-wavelength information to investigate different issues on different types of targets, even if with non-simultaneous observations. In the following paragraphs, I~will present the first results of CARMENES and GIARPS, the above-mentioned simultaneous VIS-NIR~facilities. 

\subsection{The CARMENES M Dwarf Survey} \label{sec:mdwarf}
The science operations of the CARMENES GTO have been running since January 2016, and several works have been presented, both on the spectroscopic characterization of the sample \cite{2018A&A...612A..49R,2018arXiv180110372F,2018arXiv180202946P} and on the analysis of the first RV time series.
In \cite{2018arXiv180302338T}, the first RV release of the VIS channel is provided. 
The mean RV precision of 287 targets out of 324 (in that work, only the stars with at least five RV epochs were considered) is 1.7 m s$^{-1}$, in agreement with the expectations. For targets with a spectral type later than M6, the mean RV scatter is typically larger ($\sim$10 m s$^{-1}$ with respect to the mean value of the whole sample, 3--4 m s$^{-1}$) as the $v \sin i$ increases. 
An interesting result is the null-correlation between the RVs and the activity indicators for 50\% of the considered sample. This could be explained by complex patterns of the dark spots on the stellar surface or related to the magnetic fields associated with those inhomogeneities or even a possible larger temperature contrast between the photosphere and the spots.

The first planetary companion detected by CARMENES is orbiting around the M0 star HD\,147379~
\cite{2018A&A...609L...5R}, located at 10.7 pc. This discovery has been presented by analyzing the VIS RV only, since the typical RV uncertainty reached in the NIR channel (8.6 m s$^{-1}$) is not suitable to recover this kind of signal, because of the paucity of spectral lines of early M dwarfs in that wavelength range. Therefore, the VIS RV semi-amplitude is $\sim$5 m s$^{-1}$, providing a minimum mass of $\sim$25 M$_{\oplus}$ (1.5-times the mass of Neptune). The orbital period is 88 days, which placed this planet inside the temperate zone around the star.

A test of the actual performance and capability of the instrument has been presented by~\cite{2018A&A...609A.117T}, aiming to recover a number of known rocky planets around a small sample of seven M dwarfs (single-planet systems: GJ\,15A, GJ\,176, GJ\,436, GJ\,536, GJ\,1148; multi-planet systems: GJ\,581, GJ\,876) with RV from the CARMENES VIS channel. In this work, the orbital parameters of the known systems are updated, and a new Neptune-mass companion with a period larger than 500 days in an eccentric orbit has been detected around GJ\,1148. Anyway, the more intriguing result is the non-detection of the super-Earth ($m \sin i$ = 5.35 M$_{\oplus}$) orbiting GJ\,15A with a period of 11.44 days, previously claimed with HIRES data \cite{2014ApJ...794...51H}. According to the data obtained during the first 16 months of monitoring with CARMENES ($\sim$100 RVs used), no evidence of such a signal is shown (Figure \ref{fig:gj15a}, left panel). 
This fact is corroborated by the analysis of the combined HIRES and CARMENES datasets (spanning more than 7300 days): the authors conclude that this periodicity is caused by the stellar activity since the power of the generalized Lomb--Scargle periodogram (GLS)\cite{2009A&A...496..577Z} of the corresponding peak decreases through the years. 
 \begin{figure} [H]
 \begin{center}
 \begin{tabular}{c} 
 \includegraphics[height=6.5cm,angle=0]{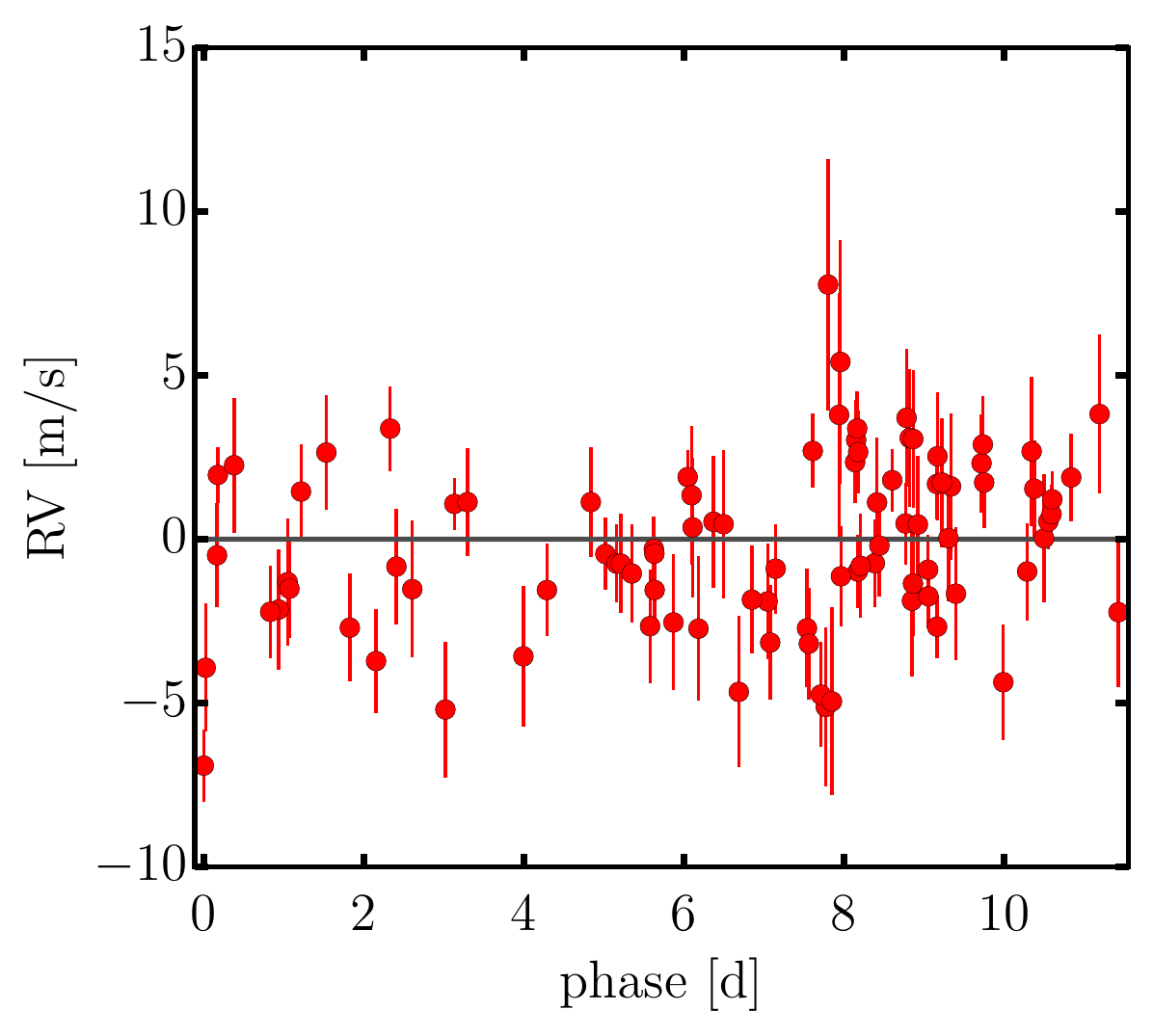}\\
 \includegraphics[height=6cm,angle=0]{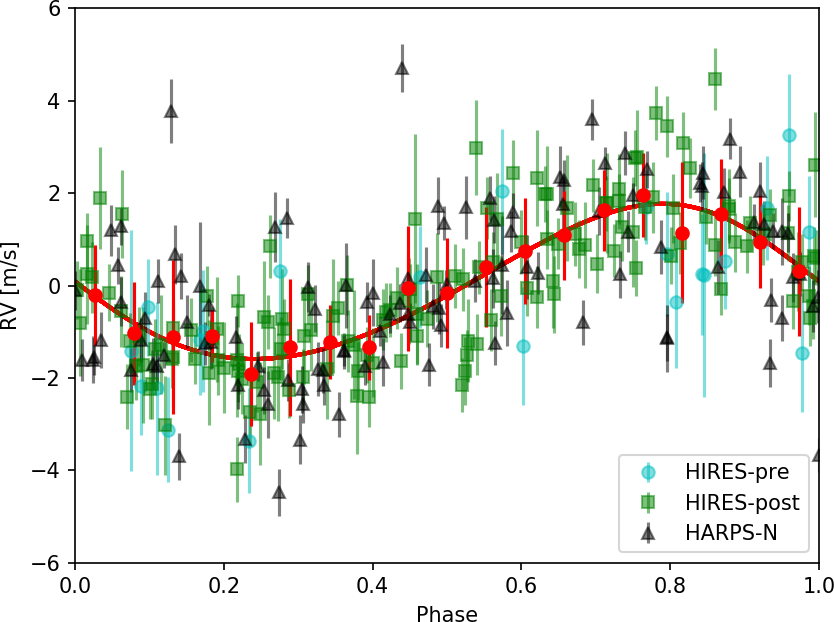}
	\end{tabular}
	\end{center}
	\vspace{-12pt}
 \caption[example] 
 { \label{fig:gj15a} The debate about GJ\,15A b. Upper panel: the VIS RVs from CARMENES showing no signal at the expected orbital period (Credit: \cite{2018A&A...609A.117T} reproduced with permission \copyright ESO). Lower panel: the RV modulation recovered from HARPS-N and HIRES RVs (Credit: \cite{2018arXiv180403476P} reproduced with permission~\copyright ESO).}
 \end{figure} 
 \vspace{-6pt}
 
However, the presence of GJ\,15Ab has been subsequently confirmed in a recent investigation of HARPS-N + HIRES data \cite{2018arXiv180403476P}, also containing the detection of a further long period planet (P~$\sim$7600~days, suggested, but not confirmed by \cite{2018A&A...609A.117T}). In \cite{2018arXiv180403476P}, the simultaneous modeling of the activity and planetary signals is performed through Gaussian processes, which retrieve the debated Super Earth (Figure \ref{fig:gj15a}, right panel), even if with a smaller value of the RV semi-amplitude with respect the previous finding \cite{2014ApJ...794...51H}, and therefore of the planet mass. According to the authors, this could be explained with a more suitable sampling of the periodicity, a higher RV accuracy of the analyzed dataset (new release of the HIRES data \cite{2017AJ....153..208B} and HARPS-N) and a better treatment of the activity contribution. Similar arguments are proposed to explain the non-detection from CARMENES data. Actually, the fitted RV semi-amplitude $K_b=1.68^{+0.17}_{-0.18}$ m s$^{-1}$ appears to be beyond the detectability limit of CARMENES, since its typical RV uncertainty is about 1.7 m s$^{-1}$ with respect to the 0.62~m~s$^{-1}$ for HARPS-N. 
A further investigation is certainly required to define which analysis is better at interpreting the observed data. The computation of the RVs in different spectral regions is proposed \cite{2018arXiv180403476P} as an additional test to evaluate the behavior of the RVs as a function of the color, {following the approach proposed by \cite{2017AJ....154..135F}. This could be done for the HARPS-N dataset since it should not be affected by a severe increase of the RV uncertainties when only a number of spectral orders are used to evaluate the RV measurement. In the case of CARMENES data, this could return a non-fully-reliable test. On the other hand,} it would be very interesting to see the contribution of the NIR RVs of CARMENES. 

CARMENES also performed an RV follow-up of the target K2-18b, a transiting Super Earth orbiting an M2.5 dwarf star, revealed by K2 photometry \cite{2015ApJ...809...25M} ({K2 is the update of the {Kepler} mission after its severe failure occurred in 2013}) and then confirmed with the NASA Spitzer observations \cite{2017ApJ...834..187B} ({\url{http://www.spitzer.caltech.edu/}}). Transit analysis returned a radius of about 2 R$_{\oplus}$. By using the VIS RVs of CARMENES ($K_b=3.5$ m s$^{-1}$), a mass of 9 M$_{\oplus}$ has been measured \cite{2018AJ....155..257S}. This implies a Super Earth having a density consistent with a solid core and a volatile-rich envelope in the temperate zone of an early M dwarf, receiving thus a similar stellar irradiation as the Earth. As a consequence, this target is a perfect candidate for the atmospheric characterization of a rocky planet in the HZ of its host star with JWST. Finally, CARMENES data tend to exclude the presence of a shorter period planet claimed with HARPS-N data \cite{2017A&A...608A..35C}, finding a signal dependent on time and wavelength. As the GLS periodograms in Figure \ref{fig:k2-18} show, while the periodicity of the 33-day planet (0.3 1/d in the frequency domain) is clearly visible in both the first and the second halves of the data, the one at nine days (0.11 1/d) vanishes in the second part of the RV monitoring. This behavior is typical when the periodicity is due to the stellar activity and not to a Keplerian signal.
 \begin{figure} [H]
 \begin{center}
 \begin{tabular}{c} 
 \includegraphics[height=4.5cm,angle=0]{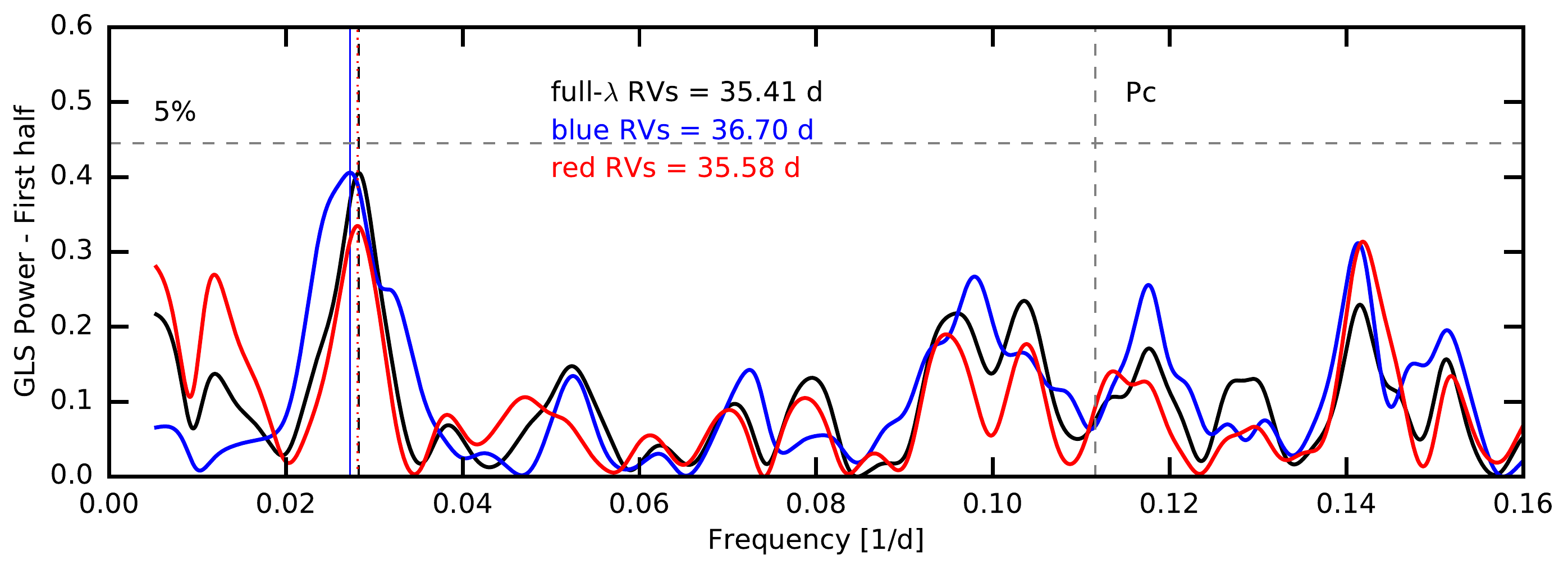}\\
 \includegraphics[height=4.5cm,angle=0]{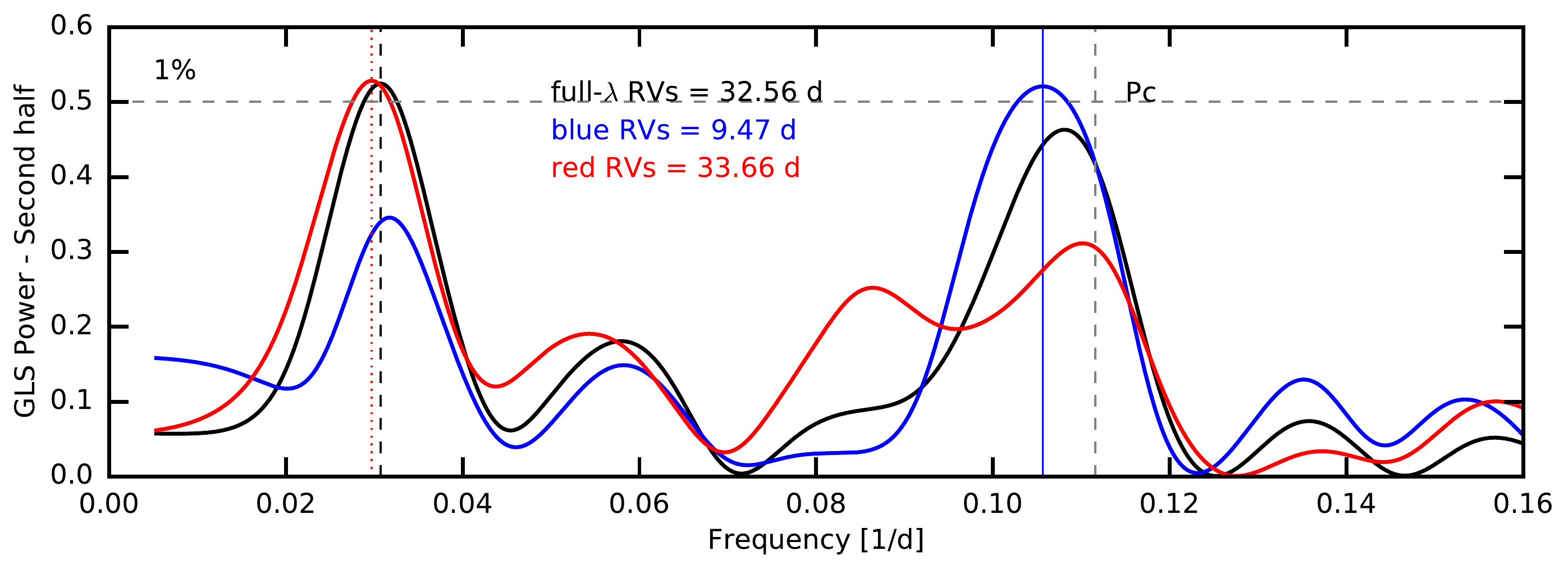}
	\end{tabular}
	\end{center}
	\vspace{-12pt}
 \caption[example] 
 { \label{fig:k2-18} GLS periodograms of the CARMENES RVs for the target K2-18b corresponding to the first half (upper panel) and the second half (lower panel) of observations. The Keplerian signal at 0.3/d is confirmed, while the one at 0.11/d is probably due to activity. Credit: \cite{2018AJ....155..257S} \copyright AAS. Reproduced with permission.}
 \end{figure} 

\subsection{Retreat of the Debated Hot Jupiter BD+20~1790 b} \label{sec:bd}

The first result obtained by using a combination of data from HARPS-N, GIANO, GIANO-B, GIARPS and the contribution of the NIR high-resolution echelle spectrograph IGRINS (see Section~\ref{gmt}) is the retreat of a controversial exoplanet around the young K5 star, BD+20\,1790 (considered as a member of the AB Dor moving group, $\sim$150 Myrs) by \cite{2018A&A...613A..50C}.
This target shows an intense and peculiar stellar activity which the responsible for the significant RV modulation in the VIS range reported by \cite{HO2010,HO2015}. This modulation shows an RV semi-amplitude K $\simeq$ 1 km s$^{-1}$, which was interpreted as a signature of a massive hot Jupiter with a period of 7.8 days. In 2010, some doubts about the real presence of this companion were raised \cite{2010A&A...513L...8F}, because the modulation was not compatible with other spectroscopic data obtained in the VIS range in a different period, thus attributing the RV variations to photospheric processes. 
The new NIR RV dataset was obtained by using telluric lines from standard stars as a reference, following the approach by \cite{2016ExA....41..351C}. 
The comparison between the VIS archive data of BD+20\,1790 and the GIANO/GIANO-B and IGRINS spectrographs (Figure~\ref{fig:bd}) shows that the average RV amplitude in the NIR is significantly lower than the VIS data (about 1/4). Moreover,~the~comparison between the previous optical data and the three RV measurements obtained with HARPS-N in the GIARPS configuration (black asterisks in the lower panel of Figure \ref{fig:bd}) shows that even the VIS modulation has changed with time, suggesting that it is caused by stellar activity. 
 \begin{figure} [H]
 \begin{center}
 \begin{tabular}{c} 
 \includegraphics[height=14cm,angle=270]{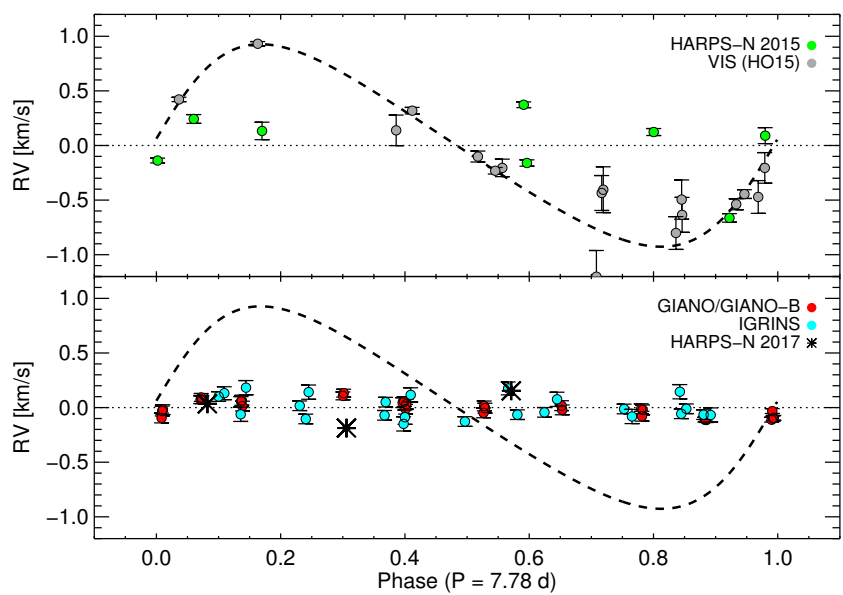}
	\end{tabular}
	\end{center}
	\vspace{-12pt}
 \caption[example] 
 { \label{fig:bd} 
{Top panel}: Orbital fit (black dashed line) obtained with the visible data by \cite{HO2010,HO2015} (grey dots) and HARPS-N 2015 RVs (green dots), overplotted. {Bottom panel}:~Orbital fit (black dashed line), GIANO/GIANO-B (red dots), IGRINS (light blue dots) and~HARPS-N 2017 (black asterisks, two acquired in GIARPS mode) RVs. Credit:  \cite{2018A&A...613A..50C} reproduced with permission \copyright ESO.}
 \end{figure} 
 \vspace{-6pt}
To have a more comprehensive indication of the photospheric activity, during the observing run of GIANO, a quasi-simultaneous photometric monitoring of BD+20\,1790 had been performed with REMIR (REM InfraRed) and ROSS2 (REM Optical Slitless Spectrograph 2), two imagers in the NIR and VIS range, respectively, mounted at the REM telescope (Rapid Eye Mount, a 60-cm robotic telescope at the ESO-La Silla Observatory). ROSS2 and REMIR can operate simultaneously thanks to a dichroic. 
Surprisingly, the VIS light curve from ROSS2 shows a smaller amplitude with respect to the NIR one obtained with REMIR (see Figure 5 in~\cite{2018A&A...613A..50C}), which is not an intuitive result, according to the arguments presented in this review. Moreover, the two modulations are in anti-phase with each other. The authors propose a photometric model that includes a mixture of cool and hot spots in the same active region able to explain the peculiarity of this star. However,~a~similar behavior has been reported by \cite{2011ApJ...736..123M,2017ApJ...836..200G} in the RVs of young stars (Section~\ref{pioneer}).

This work also demonstrates that the support of simultaneous photometry can be particularly useful both to interpret the RV curves and to add further information on the characteristics of the stellar activity.

\section{Conclusions and Perspectives}
The new scientific questions in the framework of the exoplanet search and characterization can be summarized as follow: ({i}) the search for Earth-mass rocky planets in the HZ of M dwarfs; ({ii}) the identification of the origin of planetary system diversity through the detection of planets around young stars; ({iii}) the characterization of the hot gas giant planets atmospheres as a laboratory for the future characterization of rocky habitable planets with ELTs or space-based telescopes.
The simultaneous multi-band observations in the VIS and NIR ranges are expected to be the forthcoming parameter space for these new issues. The full monitoring of stellar activity represents the key point in the detection of exoplanets around targets with enhanced and peculiar activity such as the ones considered in ({i}) and ({ii}). On the other hand, the opportunity to explore molecular features of the planetary atmospheres, present over a wide interval of wavelengths, is now possible thanks to instruments covering the range between the bluer part of the VIS up to the edge of the NIR ({iii}). 
Instruments like CARMENES, GIARPS and the forthcoming NIRPS + HARPS configuration are expected to yield a significant contribution in these and in many other fields of astrophysics. 
They also represent the starting point for future generation spectrographs that are going to equip new astronomical facilities.

{If we observe a planetary (period-mass) diagram} (or, alternatively, a (semi-major axis-mass) diagram, e.g., Figure 1 in \cite{2013Sci...340..577S}) comparing the properties of the known exoplanets with the planets of our Solar System, it is clear that the current technology is still not sufficiently adequate to find planets like our Venus, Mars, Neptune, etc., or the Earth itself. The need to reach the 1-m s$^{-1}$ (and less) RV accuracy is driven by the goal to find an Earth-like planet around a Sun-like star. 
As mentioned in this review, there are several impact factors on the achievable RV precision with high-resolution spectroscopy (both VIS and NIR): changes in the environmental conditions require a series of technical devices ensuring a high-level of instrumental stability; the wavelength calibration must rely on a very stable and reliable wavelength reference to reach the required precision; RV extraction methods must be more and more sophisticated. 
Other important technical elements can be the availability of state-of-the-art detectors and devices that allow a uniform illumination of the spectrograph slit. Last, but not least, from the observational point of view, the capability to conduct a proper data sampling of the RV signal helps to avoid the aliasing phenomenon that hampers the frequency (period) analysis.
We are quite confident to say that all of these issues have been successfully addressed, at least in the visible range, at least to obtain the 1-m s$^{-1}$ regime. As previously stated, the technology transfer from VIS to NIR is far from immediate, but many efforts have been already made to allow precise RV also in the NIR domain, starting from the lessons learned in the last few decades.

At this moment, we are approaching the new milestone of the Doppler method, since we are waiting for the precise NIR RV measurement from CARMENES (1 m s$^{-1}$ expected), on the one hand, and the outstanding RV precision of 10 cm s$^{-1}$ from ESPRESSO, on the other. On the scientific side, the challenge is to be able to manage all the other detectable effects, first of all, the stellar activity. Even in this case, valuable examples of data treatment are presented in literature, and it will be very interesting to see how they will work in the new ultra-precise RV domain.
Reaching the cm s$^{-1}$ regime is a necessary step of the journey, but it is also crucial to know how to exploit this opportunity.
In the near future, these new facilities will necessarily be more and more specialized for RV monitoring for the search for exoplanets (some of them are already employed in such a way); just think about the follow-up of the forthcoming space satellite dedicated to transit detection. 
Actually,~the~cooperation among different detection methods (and thus, instruments and projects) will probably be the best way to characterize a star-planet system. In this context, an important role will be played by all the tools that allow obtaining a more precise estimate of the stellar parameters like mass and radius (e.g.,~stellar~modeling and asteroseismology), since they are crucial to constrain the global planet properties.


\vspace{6pt} 


\funding{This research received no external funding.}

\acknowledgments{S.B. is grateful to A. F. Lanza, I. Carleo and R. Claudi for the useful comments on this manuscript and the support with the specific topics, to C. Dressing for her kind advice and to M. Endl for the update on HRS-2. S.B. also wishes to thank the two anonymous referees that provided valid and useful suggestions that allowed this paper to be more complete and accurate. Finally, S.B. is grateful to the authors and the journal editorial boards for the permission to reproduce their plots and figures in this review.}


\conflictsofinterest{As Project Manager of the GIARPS instrument installed at the INAF TNG (Telescopio Nazionale Galileo), the author was deeply involved in its technical and scientific development, in the commissioning phase and science exploitation of the data. Nevertheless, in this manuscript, she tried to be as impartial as possible, just reporting the main characteristics and interesting results of both GIARPS and CARMENES instruments now available, as well as the forthcoming projects. The only aim of this paper is to show the outstanding potential of the multi-wavelength observations in the framework of exoplanet detection and characterization. No founding sponsors had a role in the design of the study; in the collection, analyses or~interpretation of data; in the writing of the manuscript; and in the decision to publish the results.
}

\abbreviations{The following abbreviations are used in this manuscript:\\

\noindent 
\begin{tabular}{@{}ll}
AO & Adaptive Optic \\
CARMENES & \makecell[l]{Calar Alto high-Resolution search for M dwarfs with \\ Exo-Earths with Near-infrared and optical \'Echelle Spectrographs} \\
CRIRES & Cryogenic high-resolution infrared echelle spectrograph \\
E-ELT & European Extremely Large Telescope \\
ELT & Extremely Large Telescope \\
ESO & European Southern Observatory \\
FP & Fabry--Perot \\
G-CLEF & GMT Consortium Large Earth Finder \\ 
GLS & Generalized Lomb--Scargle \\
GMT & Giant Magellan Telescope \\
GMTNIRS & GMT Near-IR spectrograph \\
HARPS & High Accuracy Radial velocity Planet Searcher \\
HIRES & High Resolution Spectrograph \\
HZ & habitable zone \\
IGRINS & Immersion Grating Infrared Spectrometer \\
JWST & James Webb Space Telescope \\
INAF & (Italian) National Institute for Astrophysics \\
NIR & Near-infrared \\
NIRPS & Near Infra Red Planet Searcher \\
RV & radial velocity \\
TNG & Telescopio Nazionale Galileo \\
VIS & visible \\
VLT & Very Large Telescope \\
\end{tabular}
}

\appendixtitles{yes} 
\appendixsections{multiple} 
\appendix
\section{Other VIS and NIR Facilities}\label{sec:app}
In this Appendix, a list of new high-resolution spectrographs in the VIS or NIR range (i.e., without the other counterpart) providing high-precision RV in public facilities is reported. Table \ref{app} is an excerpt from \cite{2017RNAAS...1...51W}, where a more detailed description of these instruments is available. Instruments already discussed in dedicated sections of this manuscript are not included.
A link to the dedicated web pages of each spectrograph are provided, with the aim to be as updated as possible about the technical/scientific progress. 

\clearpage
\begin{table}[H]
\begin{center}
\begin{tabular}{l l c p{5cm}}\toprule
 \multirow{2}{*}{\textbf{Spectrograph}} 
 &\multirow{2}{*}{\textbf{Telescope and Site}} & \textbf{Expected} &\multirow{2}{*}{\textbf{Link}} \\
 & & \textbf{Availability}&\\
\midrule
\textit{Visible:} & & & \\ \hline
ESPRESSO  & VLT (4 $\times$ 8-m, Paranal, Chile) & 2018 & \url{https://www.eso.org/sci/facilities/paranal/instruments/espresso.html} \\ \hline
PEPSI  & LBT (8.4-m, Mt. Graham, USA) & 2018 & \url{https://pepsi.aip.de/} \\ \hline
HRS-2 & HET (10-m, McDonald, USA) & 2020 & \url{https://hydra.as.utexas.edu/?a=help&h=93} \\ \hline
MAROON-X & Gemini N. (8.1-m, Maunakea, USA) & 2019 & \url{https://www.gemini.edu/sciops/instruments/maroon-x} \\ \hline
NEID & WIYN (3.5-m, Kitt Peak, USA) & 2019 & \url{http://neid.psu.edu/} \\ \hline
Veloce & AAT (4-m, Siding Spring, UAS) & 2020 * & \url{https://www.aao.gov.au/science/instruments/current/veloce/overview} \\ \hline
HARPS3 & INT (2.5-m, La Palma, Spain) & 2020 & \url{http://www.terrahunting.org/harps3.html} \\ \hline
KPF & Keck I (10-m, Maunakea, USA) & 2020 & \cite{2016SPIE.9908E..70G}, \url{https://www2.keck.hawaii.edu/inst/kpf/} \\
\hline
\textit{Near Infrared:} & & & \\ \hline
iSHELL & IRTF (3-m, Maunakea, USA) & 2016 & \url{http://irtfweb.ifa.hawaii.edu/~ishell/} \\ \hline
HPF & HET (10-m, McDonald, USA) & 2018 & \url{https://hpf.psu.edu/} \\ \hline
IRD & Subaru (8.2-m, Maunakea, USA) & 2018 & \url{http://ird.mtk.nao.ac.jp/IRDpub/index_tmp.html} \\ \hline
PARVI & Hale (5.1-m, Mt. Palomar, USA) & 2019 & \url{https://techport.nasa.gov/view/92844} \\ \hline
SPIRou & CFHT (3.6-m, Maunakea, USA) & 2019 & \url{http://www.cfht.hawaii.edu/en/projects/SPIRou/} \\
\bottomrule
\end{tabular}
\begin{tabular}{ccc}
\multicolumn{1}{c}{\footnotesize * The red channel should be available by the end of 2018.}
\end{tabular}
\end{center}
\bigskip
\caption{List of high-resolution spectrographs in the VIS or NIR range recently integrated and commissioned, or close to their first light.}  \label{app}
\end{table}

\vspace{-36pt}



\reftitle{References}


\end{document}